\documentstyle[prb,aps,eqsecnum,twocolumn]{revtex}
\input BoxedEPS.tex
\SetTexturesEPSFSpecial
\HideDisplacementBoxes
\begin{document}
\draft
\twocolumn[\hsize\textwidth\columnwidth\hsize\csname@twocolumnfalse\endcsname
%
\title{ Theory of itinerant-electron ferromagnetism }
%
\author{Fusayoshi J. Ohkawa}
\address{Division of Physics, Graduate School of Science, 
Hokkaido University, Sapporo 060-0810, Japan}
\date{Received: \hspace{4cm} }
\maketitle
\begin{abstract} 
A theory of Kondo lattices or a $1/d$ expansion theory, with
$d$ spatial dimensionality, is applied to studying
itinerant-electron ferromagnetism. Two relevant multi-band
models are examined: a band-edge model where the chemical
potential is at one of band-edges, the top or bottom of 
bands, and a flat-band model where one of bands is almost
flat or dispersionless and the chemical potential is at the
flat band. In both the models, a novel ferromagnetic exchange
interaction arises from the virtual exchange of pair
excitations of quasiparticles; it has two novel properties such
as its strength is in proportion to the effective Fermi energy
of quasiparticles and its temperature dependence is responsible
for the Curie-Weiss law.  When the Hund coupling $J$ is strong
enough, the superexchange interaction, which arises from the
virtual exchange of pair excitations of electrons across the
Mott-Hubbard gap, is ferromagnetic. In particular, it is
definitely ferromagnetic for any nonzero $J>0$ in the large
limit of band multiplicity. Ferromagnetic instability occurs,
when the sum of the two exchange  interactions is ferromagnetic
and it overcomes the quenching of magnetic moments by the Kondo
effect or local quantum spin fluctuations and  the suppression
of magnetic instability by the mode-mode coupling among
intersite spin fluctuations.
\end{abstract}
\vspace{0.0cm}~\    ]

\narrowtext

\section{Introduction}\label{SecIntroduction}

Itinerant-electron ferromagnetism has been a long standing and
important issue since the advent of quantum mechanics. When the
Hubbard model, for example, is considered in the Hartree-Fock
approximation,  a ferromagnetic instability condition is given
by
\begin{equation}\label{EqStrongU}
U\rho_0(0)>1 ,
\end{equation}
with $U$ the on-site repulsion and $\rho_0(0)$ the
unrenormalized density of states at the chemical potential.
Local correlations between electrons with parallel spins are
accurately treated in any approximation as long as the Pauli
principle is taken into account. In the Hartree-Fock
approximation, however, no local correlations between electrons
with antiparallel spins are taken into account; paramagnetic
states are disfavored so that magnetic states are relatively
favored. The condition (\ref{EqStrongU}) is never reliable.
Local spin correlations should be accurately taken into account
far beyond the Hartree-Fock approximation in any reliable
theory.

In 1963, Kanamori\cite{Kanamori} studied ferromagnetism in Ni
in the $T$-matrix approximation, which is relevant for low
densities of electrons or holes. He showed that a ferromagnetic
state is stabilized not only when Eq.~(\ref{EqStrongU}) is
satisfied but also when the density of states has a sharp peak
within broad bands and the chemical potential is at the peak.
Mielke\cite{Mielke1}, Mielke and Tasaki\cite{Mielke2},
Tasaki\cite{Tasaki}, and Kusakabe and Aoki\cite{Kusakabe}
showed  that the ground state  of certain multi-band models is
ferromagnetic.  In their models, one of bands is 
dispersionless or flat and the density of states is similar to
that of Kanamori's. These theories show that the existence of
a flat band or flat bands is one of the most relevant
conditions for itinerant-electron ferromagnetism.  In many 
ferromagnets such as Ni, the density of electrons or holes is
low and the chemical potential is at one of the band edges. One
may argue that this condition must also be relevant. Call these
two conditions flat-band and band-edge conditions in this
paper.

The ground state can be ferromagnetic even if neither of these
conditions is satisfied. Nagaoka\cite{Nagaoka} and
Thouless\cite{Thouless} showed that when a single {\it hole} is
introduced into the just half-filled system or the electron
number is $N-1$, with $N$ the number of lattice sites, the
ground state is a completely polarized ferromagnetic state in
the limit of $U\rho_0(0) \rightarrow +\infty$. Because spin
waves of this state are abnormal and any extension of their
arguments to low but nonzero densities of {\it holes} is
difficult, it is doubtful if a ferromagnetic state is
stabilized in the thermodynamic limit of small {\it hole}
densities where the number of electrons is $N(1-\delta)$,
with $N\rightarrow +\infty$ and  $\delta \rightarrow +0$.

In 1963,  another two  distinguished
papers\cite{Hubbard,Gutzwiller} were published.
Hubbard\cite{Hubbard} showed that a band splits into two bands,
the lower and upper Hubbard bands. When the density of states of
unrenormalized electrons is a Lorentzian shape of width 
$\Gamma\simeq 1/\pi\rho_0(0)$, for example, the local part of
the Green function is given by 
\begin{eqnarray}\label{EqHubTh}
\frac{1}{N} \! \sum_{\bf k}\!
G_{\sigma}(i \varepsilon_{n}, {\bf k}) 
&=&
\frac{1}{i \varepsilon_{n} \!+\! \mu \!-\! \epsilon_{a} \!-\!
\tilde{\Sigma}_{\sigma}(i \varepsilon_{n} ) 
\!+\! i \Gamma \mbox{sgn}(\varepsilon_{n}) }
\nonumber \\ &=&
\frac{1-\langle n_{-\sigma}\rangle}
{i\varepsilon_{n}\!+\!\mu\!-\!\epsilon_{a}\!+\!
i\Gamma \mbox{sgn}(\varepsilon_{n})} 
\nonumber \\ &&  + 
\frac{\langle n_{-\sigma}\rangle}
{i\varepsilon_{n}\!+\!\mu\!-\!\epsilon_{a}\!-\!
U\!+\!i\Gamma\mbox{sgn}(\varepsilon_{n})} ,
\end{eqnarray}  
with  
$\tilde{\Sigma}_{\sigma}(i\varepsilon_{n} )$ the selfenergy,
$\mu$ the chemical potential, $\epsilon_{a}$ the band center,
and $\langle n_{\sigma}\rangle$ the average number of electrons
with spin $\sigma$ per site.  When the Fermi-liquid
relation\cite{Luttinger1,Luttinger2} is made use of,
Gutzwiller's result\cite{Gutzwiller} gives 
the mass enhancement factor of 
\begin{equation}\label{EqGutPhi}
\tilde{\phi}_\gamma \simeq 1/(1-n) ,
\end{equation} 
with $n=\langle n_{\uparrow}\rangle+\langle
n_{\downarrow}\rangle$, so that the quasi-particle bandwidth
is approximately given by $1/\tilde{\phi}_\gamma\rho_0(0)$.
Hubbard's result is relevant for single-particle excitations
far from the chemical potential, while Gutzwiller's
result is relevant for those in the vicinity of the chemical
potential. A combined theory of Hubbard's and Gutzwiller's
gives a three-peak structure of the density of states,
Gutzwiller's band at the chemical potential between the lower
and upper Hubbard bands.

The selfenergy obtained by Hubbard does not depend on
wave-numbers {\bf k} and the mass enhancement factor obtained
by Gutzwiller neither depends on {\bf k}. The two theories are
within the single-site approximation (SSA).

Treating the Hubbard or the periodic Anderson model in one of
the best SSA's that include all the single-site terms is
reduced to determining and solving selfconsistently a
single-impurity Anderson model.\cite{Mapping}  The Anderson
model is one of the simplest effective Hamiltonians for the
Kondo problem. The Kondo problem is almost completely
understood;\cite{Wilson,Nozieres,Yamada,Yosida} even the exact
solution by the Bethe method was obtained.\cite{Exact} Then,
many useful results are available to clarify single-site
properties in lattice systems. The most essential physics
involved in the Kondo problem is the quenching of magnetic
moments by local quantum spin fluctuations, so that the ground
state is singlet and nonmagnetic\cite{YosidaSing} although it
is close to a magnetic quantum critical point. No magnetic
instability occurs within the SSA, and it  can occur only when
intersite or multi-site effects are properly considered.
Multi-site effects can be perturbatively considered by starting
from an {\it unperturbed} state constructed in the SSA. Because
the quenching of magnetic moments by local quantum spin
fluctuations is properly treated in such a perturbative theory,
it must give a reliable magnetic instability condition. Such a
perturbative theory is nothing but a theory of Kondo lattices. 
It is also nothing but a $1/d$ expansion theory, with $d$ being
spatial dimensionality.  Any SSA that includes all the
single-site terms is rigorous for Landau's Fermi-liquid states
in infinite dimensions $(d\rightarrow+\infty)$.\cite{Metzner}
Multi-site  terms are of higher order in $1/d$ except for
Weiss' mean fields, which are responsible for magnetic
instability.

The spin susceptibility of magnets obeys the Curie-Weiss law.
Two mechanisms have been proposed for that of itinerant-electron
magnets:\cite{ComCW} the suppression of the spin
susceptibility  by the mode-mode coupling among intersite spin
fluctuations\cite{Murata,MK,Moriya} and its enhancement by
Weiss' mean fields.\cite{Ohkawa-CW,Miyai}
When the density of states is almost constant in the vicinity
of the chemical potential, the suppression by the mode-mode
coupling becomes larger with increasing temperatures; it gives
the Curie-Weiss law.\cite{Murata,MK,Moriya}  When the density
of states has a sharp peak at the chemical potential, on the
other hand, the suppression becomes larger with decreasing
temperatures; the mode-mode coupling plays a totally negative
role in the Curie-Weiss law in the flat-band
model.\cite{Miyai}

Because the mode-mode coupling suppresses the spin
susceptibility in any case, it cannot be a mechanism of
magnetic instability.  A scenario that the mechanism of the
Curie-Weiss law is also responsible for  magnetic instability
itself seems to be more reasonable than the scenario that the
mechanism of the Curie-Weiss law suppresses magnetic
instability.

The mode-mode coupling is of higher order in $1/d$ so that it
vanishes in the large limit of $d$.  There must be another
mechanism, which is of leading order in $1/d$. Only the
possible one is the temperature dependence of Weiss' mean
fields.\cite{ComCW} The enhancement of the spin susceptibility 
by an exchange interaction arising from the virtual exchange of
pair excitations of quasi-particles can give the Curie-Weiss
law.\cite{Ohkawa-CW}  Polarizations multiplied by the exchange
interaction are Weiss' mean fields, and they are of leading
order in $1/d$. When the flat-band or band-edge conditions is
satisfied, this exchange interaction is ferromagnetic and the
Curie-Weiss law appears only in the homogeneous
susceptibility.  Miyai and the present author\cite{Miyai}
examined the Curie-Weiss law of Ni by using Kanamori's model.

Field-induced ferromagnetism or metamagnetism is observed in
many metals. In CeRu$_{2}$Si$_{2}$, for example, such a 
ferromagnetic state is stabilized at  magnetic fields higher
than $H_{M}\simeq 7.7$ T. One of the most crucial experimental
results is the single-parameter scaling;\cite{scaling,scaling2}
many quantities sensitively depend on pressures, and they scale
with the Kondo temperature or energy, $k_BT_K$. The Kondo
energy is  approximately equal to a half of the bandwidth of
quasiparticles, 
$k_BT_K \simeq 1/[2\tilde{\phi}_\gamma\rho_0(0)]$. This 
experiment gives a strong restriction for a ferromagnetic
exchange interaction responsible for the metamagnetism in
CeRu$_{2}$Si$_{2}$.
Satoh and the present author showed that the exchange
interaction responsible for the Curie-Weiss law is also
responsible for the metamagnetism.\cite{Satoh} In 4$f$ electron
systems, the density of states is of a camelback structure or
of a two-peak structure because of the hybridization between
dispersive conduction bands and almost dispersionless $f$
bands. In the absence of magnetic fields, the chemical
potential is between the two peaks; either of the flat-band and
band-edge conditions is not satisfied, so that the exchange
interaction is antiferromagnetic. In the presence of fields as
high as $H_{M}$, the chemical potential is at one of the peaks
of the camelback because of the Zeeman shift of up and down
spin bands; both of  the flat-band and band-edge conditions
are  satisfied, so that the exchange interaction becomes
ferromagnetic around $H_{M}$. Because this exchange interaction
has a novel property that its strength is proportional to  the
bandwidth of quasiparticles $1/\tilde{\phi}_\gamma\rho(0)$, it
is easy to explain  the observed single-parameter scaling. Call
this one simply a novel exchange interaction in this paper.

The previous papers\cite{Miyai,Satoh} imply that when either or
both of the flat-band and band-edge conditions are satisfied
ferromagnetic instability is caused by  the novel exchange
interaction even in the absence of fields. However, there is a
crucial drawback in the previous  papers.\cite{Miyai,Satoh}  An
additional phenomenological ferromagnetic exchange interaction,
which is independent of temperatures and magnetic fields, was
assumed to obtain a quantitative agreement between experiment
and theory. The authors of these papers speculated that such a
ferromagnetic exchange interaction must exist in multi-band
models.

The purpose of this paper is to study ferromagnetism in
multi-band models from a theoretical viewpoint of the
competition between magnetic instability caused by intersite
exchange interactions and the quenching of magnetic moments 
by the Kondo effect. This paper is organized as follows: In
Sec.~\ref{SecPrel}, an {\it unperturbed} state in the $1/d$
expansion is constructed in the SSA.  It is shown in
Sec.~\ref{SecExch} that when the Hund coupling is strong enough
the superexchange interaction, which arises from the virtual
exchange of pair excitations between the lower and upper
Hubbard bands, becomes ferromagnetic.  It is shown in
Sec.~\ref{SecModel} that when the flat-band or band-edge
condition is satisfied a ferromagnetic state can be
stabilized.  Discussion is given in Sec.~\ref{SecDis}, and
summary is given in Sec.~\ref{SecSum}. In
Appendix~\ref{SecApp}, a single-impurity Anderson model is
studied by the $1/(2l+1)$ expansion method, with $2l+1$ being
orbital degeneracy.

\section{Preliminaries}\label{SecPrel}
\subsection{Multi-band model with the Hund coupling}
\label{MultiModel}

Consider a $(2l+1)$-band model, with $l$ an integer or a
half integer:\cite{Com1}
\begin{equation}\label{EqH}
{\cal H} = {\cal H}_{0} + {\cal H}_{I} .
\end{equation}
The first term is a one-body term given by
\begin{equation}\label{EqH0}
{\cal H}_{0} = 
\hspace{-0pt}\sum_{mm^\prime} \sum_{{\bf k} \sigma}
\Bigl[ E_{mm^\prime}({\bf k})  - \mu\delta_{mm^\prime} \Bigr] 
a_{m{\bf k}\sigma}^\dagger a_{m^\prime{\bf k}\sigma } ,
\end{equation}
with  $a_{m{\bf k}\sigma}^\dagger$ a creation operator
of carriers, electrons or holes. Bands or orbitals are
denoted by $m= -l$, $-(l-1)$, $\cdots$, and $l$, and up and down
spins by $\sigma=\pm 1$. 
It is assumed that all the bands are degenerate at the $\Gamma$
point due to lattice symmetry, so that both $E_{mm}(0)$ and
\begin{equation}\label{EqLvA}
\epsilon_a = \frac{1}{N} \sum_{\bf k}
E_{mm}({\bf k}) 
\end{equation}
do not depend on $m$, and $E_{mm^\prime}(0)=0$ and 
$\bigl(1/\sqrt{N}\bigr)\sum_{\bf k}$ 
$E_{mm^\prime}({\bf k})X_{inv}({\bf k})\!=\!0$
for $m\ne m^\prime$ and any  function 
$X_{inv}({\bf k})$ invariant under any symmetry
transformation.
In this paper,  
$E_{mm^\prime}({\bf k})\!=\!E_{m^\prime m}({\bf k})$
being real and
\begin{equation}\label{EqSym}
E_{mm^\prime}({\bf k}) = \left\{
\begin{array}{ll}
E_{ll}({\bf k}) ,  & m=m^\prime \vspace{0.2cm}\\
E_{l-l}({\bf k}) , & m \ne m^\prime 
\end{array}\right. 
\end{equation}
are also assumed.\cite{Com2} This assumption makes
arguments in this paper quite simple, as is shown later. The
dispersion relations of
$2l+1$ bands are given by 
\begin{equation}\label{EqDisper}
E_m^*({\bf k}) = \left\{\begin{array}{l}
E_{ll}({\bf k}) + 2l E_{l-l}({\bf k})\hspace{1pt} ,\ m=l 
\vspace{0.2cm}\\ E_{ll}({\bf k}) - E_{l-l}({\bf k})\hspace{1pt}  ,\
-l \le m \le (l-1)
\end{array} \right. .
\end{equation}
The flat-band condition is satisfied when the $l$th band
is almost dispersionless in such a way that
\begin{equation}\label{EqFlat}
E_l^* ({\bf k}) = E_{ll}({\bf k}) + 2l E_{l-l}({\bf k}) \simeq
E_{ll}(0) 
\end{equation} 
for any ${\bf k}$. In this case, the dispersion relations of
other $2l$ bands $(m\ne l)$ are given by
\begin{equation}
E_m^* ({\bf k}) \simeq E_{ll}({\bf k}) + \frac{1}{2l} \bigl[
E_{ll}({\bf k})  - E_{ll}(0) \bigr] .
\end{equation}

The second term in Eq.~(\ref{EqH})
is the on-site Coulomb interaction:
\begin{mathletters}\label{EqInt}
\begin{eqnarray}\label{EqIntA}
{\cal H}_{I} &=& \frac{1}{2}\sum_{mm^\prime}
\sum_{i\sigma\sigma^\prime}
U_{mm^\prime}
a^{\dag}_{mi\sigma}a^{\dag}_{m^\prime i\sigma^\prime}
a_{m^\prime i\sigma^\prime}a_{mi\sigma} \nonumber \\
&& + \frac{1}{2} J \sum_{m\ne m^\prime}
\sum_{i\sigma\sigma^\prime}
a^{\dag}_{mi\sigma}a^{\dag}_{m^\prime i\sigma^\prime}
a_{mi\sigma^\prime}a_{m^\prime i\sigma} 
\\ \label{EqIntB}
&=& \frac{1}{2} U \sum_{mi\sigma} n_{mi\sigma}n_{mi-\sigma} 
\nonumber \\ &&
+ \frac{1}{2} \left( U^\prime - \frac{1}{2}J \right)
\sum_{m\ne m^\prime}
\sum_{i\sigma\sigma^\prime}
n_{mi\sigma}n_{m^\prime i\sigma^\prime} 
\nonumber \\ && - J \sum_{m\ne m^\prime}\sum_{iX}
\sum_{\sigma_1\sigma_2}\sum_{\sigma_{3}\sigma_{4}}
\left( s_{X}^{\sigma_1\sigma_2} s_{X}^{\sigma_3\sigma_4}\right) 
\nonumber \\ && \hspace*{1.2cm} \times 
a^{\dagger}_{mi \sigma_{1}}a_{mi \sigma_{2}}
a^{\dagger}_{m^\prime i \sigma_{3}}a_{m^\prime i \sigma_{4}} ,
\end{eqnarray}
\end{mathletters}
with
\begin{equation}
a_{mi\sigma}^{\dagger} = \frac{1}{\sqrt{N}} \sum_{\bf k} 
e^{-i ({\bf k} \cdot {\bf R}_{i} )} a_{m{\bf k}\sigma}^{\dagger} ,
\end{equation}
$a_{mi\sigma}$ its Hermite conjugate, and
%
$n_{mi\sigma} = 
a_{mi \sigma}^{\dag} a_{mi \sigma} $.
%
%
Because of the degeneracy of $2l+1$ bands at the $\Gamma$ point,
\begin{equation}\label{EqSymU}
U_{mm^{\prime}} = 
\left\{ \begin{array}{ll}
U, & m = m^\prime \vspace{0.1cm}\\
U^\prime, & m \ne m^\prime
\end{array} \right. 
\end{equation}
is also assumed. In general, the Hund coupling is ferromagnetic
$(J>0)$.
In Eq.~(\ref{EqIntB}), $s_{X}^{\sigma\sigma^\prime}$ is the 
$({\sigma\sigma^\prime})$th element of the $S=\frac{1}{2}$ spin
matrix, $s_{x}$, $s_{y}$, or $s_{z}$. In this paper, 
$s_{z}$ is diagonalized so that
$s_{z}^{\sigma\sigma^\prime}=\frac{1}{2} \sigma
\delta_{\sigma\sigma^\prime}$.

\subsection{Mapping to the Anderson model} 
%
%

The single-particle Green function for ${\cal H}$ is given by
\begin{equation}
{\cal G}_{\sigma}^{-1}(i\varepsilon_{n},{\bf k})
= 
\left( i\varepsilon_{n}+\mu \right) {\cal I} 
- {\cal E}({\bf k})
- {\cal S}_{\sigma}(i\varepsilon_{n},{\bf k}) ,
\end{equation}
with ${\cal I}$ being the unit matrix,
$\bigl[\hspace{1pt} {\cal E}({\bf k}) \bigr]_{mm^\prime} = 
E_{mm^\prime} ({\bf k})$,
and ${\cal S}_{\sigma}(i\varepsilon_{n},{\bf k}) $
the selfenergy matrix. 
The selfenergy is divided into
single-site and multisite terms: 
\begin{equation}
{\cal S}_{\sigma}(i\varepsilon_{n},{\bf k}) =
\tilde{\Sigma}_{\sigma}(i\varepsilon_{n}) {\cal I} +
\Delta {\cal S}_{\sigma}(i\varepsilon_{n},{\bf k}) ,
\end{equation}
with $\tilde{\Sigma}_{\sigma}(i\varepsilon_{n})$ 
the single-site selfenergy and 
$\Delta{\cal S}_{\sigma}(i\varepsilon_{n},{\bf k})$
the multisite one. 
In our model, the selfenergy for ${\bf k}=0$ is proportional
to the unit matrix because of the assumed symmetry;
$\bigl[\hspace{1pt} \Delta{\cal S}_{\sigma}(i\varepsilon_{n},0)
\bigr]_{mm} $
does not depend on $m$ and
$\bigl[\hspace{1pt} \Delta{\cal S}_{\sigma}(i\varepsilon_{n},0)
\bigr]_{mm^\prime} =0$
for $m\ne m^\prime$.

Calculating the single-site selfenergy 
 $\tilde{\Sigma}_{\sigma}(i\varepsilon_{n})$ 
is reduced to solving 
a single-impurity Anderson model:
\begin{eqnarray}\label{EqHamA}
{\cal H}_{A} &=& 
\sum_{m\sigma} \bigl[ 
\epsilon_{a}-\mu \bigr]n_{m\sigma} +
\sum_{{\bf k}\sigma} \bigl[e_{c}({\bf k})-\mu \bigr]
c^{\dag}_{{\bf k}\sigma}c_{{\bf k}\sigma} 
\nonumber \\ && 
+\frac{1}{\sqrt{N}} \sum_{m{\bf k} \sigma} \left[
v_{m{\bf k}}a^{\dag}_{m\sigma}c_{{\bf k}\sigma}
+v_{m{\bf k}}^{*}c^{\dag}_{{\bf k}\sigma}a_{m\sigma} \right] 
\nonumber \\ &&
+\frac{1}{2}\sum_{mm^\prime}\sum_{\sigma\sigma^\prime}
U_{mm^\prime}
a^{\dag}_{m\sigma}a^{\dag}_{m^\prime\sigma^\prime}
a_{m^\prime\sigma^\prime}a_{m\sigma} 
\nonumber \\ && + \frac{1}{2} J \sum_{m\ne m^\prime}
\sum_{\sigma\sigma^\prime}
a^{\dag}_{m\sigma}a^{\dag}_{m^\prime\sigma^\prime}
a_{m\sigma^\prime}a_{m^\prime\sigma}  ,
\end{eqnarray}
with
%
$n_{m\sigma} = a^\dag_{m\sigma}a_{m\sigma}$.
%
Here, $\epsilon_{a}$, $\mu$, $U_{mm^\prime}$, and $J$ 
are the same as those for the lattice system;
$e_{c}({\bf k})$ and $v_{m{\bf k}}$
should be determined so as to satisfy
a mapping condition discussed below.  
%
%
The Green function for ${\cal H}_{A}$ is given by
\begin{equation}
\tilde{\cal G}_{\sigma}^{-1}(i\varepsilon_{n}) =
\left[i\varepsilon_{n} + \mu 
- \tilde{\Sigma}_{\sigma}(i\varepsilon_{n}) 
- \tilde{L}_{\sigma}(i\varepsilon_{n}) \right] {\cal I},
\end{equation}
with
%
$\tilde{L}_{\sigma}(i\varepsilon_{n}) = 
(1/N) \sum_{\bf k} |v_{m{\bf k}}|^{2} \big/
\bigl[i\varepsilon_{n}-e_{c}({\bf k})\bigr] $.
%
Because of the assumed symmetry, 
$\tilde{\cal G}_{\sigma}^{-1}(i\varepsilon_{n})$ is diagonal
and $\tilde{L}_{\sigma}(i\varepsilon_{n})$ does not depend on
$m$.
Because the symmetrical property of
$\bigl[ \Delta{\cal S}_{\sigma}(i\varepsilon_{n},{\bf k})
\bigr]_{mm^\prime} $
is the same as that of $E_{mm^\prime}({\bf k})$,
the local Green function defined by 
\begin{equation}
{\cal R}_{\sigma}(i\varepsilon_{n}) =
\frac{1}{N}\sum_{\bf k} {\cal G}_{\sigma}
(i\varepsilon_{n},{\bf k}) 
\end{equation}
is also diagonal. When 
\begin{equation}\label{EqMap}
\tilde{\cal G}_{\sigma}(i\varepsilon_{n}) = 
{\cal R}_{\sigma}(i\varepsilon_{n}) 
\end{equation}
is satisfied any single-site quantity of the lattice system is
the same as its corresponding quantity of  the single-impurity
Anderson model. The quantities, $\epsilon_{c}({\bf k})$ and
$v_{m{\bf k}}$, of the Anderson model  should be
selfconsistently determined to satisfy Eq.~(\ref{EqMap}); 
Eq.~(\ref{EqMap}) is nothing but the mapping condition  to the
Anderson model.

\section{Exchange interactions}\label{SecExch}
\subsection{Kondo lattices}\label{SecExchDef} 

Consider the mapped Anderson model (\ref{EqHamA}), and introduce
fictitious and infinitesimally small external fields such as
%
${\cal H}_{ext}=-\sum_{m\sigma}\Delta E_{m\sigma}n_{m\sigma}$,
%
with
\begin{equation}\label{EqExt}
\Delta E_{m\sigma} = \Delta\mu + 
\mu_B \! \left[\hspace{-1pt} \frac{1}{2} \sigma g_s H_s  
\!+\! m g_o H_o \!+\! \frac{1}{2} 
(\hspace{-1pt}\sigma m) g_{so}H_{so}
\right].
\end{equation}
Here, $\mu_B$ is the Bohr magneton; $g_s$, $g_o$ and
$g_{so}$ are $g$ factors.\cite{ComLargeL}  The single-site
selfenergy at $T=0~$K is expanded in such a way that
\begin{eqnarray}\label{EqSelf}
\tilde{\Sigma}_{m\sigma}(i \varepsilon_{n}) &=&
\tilde{\Sigma}(0) 
+ ( 1 - \tilde{\phi}_\gamma) i \varepsilon_{n} 
+ ( 1 - \tilde{\phi}_{c}) \Delta\mu
\nonumber \\ && 
+ ( 1 - \tilde{\phi}_{s}) \frac{1}{2}\sigma g_s\mu_B H_s
+ ( 1 - \tilde{\phi}_{l}) m g_o \mu_B H_o
\nonumber \\ && 
+ ( 1 - \tilde{\phi}_{so}) \frac{1}{2}(\sigma m) g_{so}
\mu_B H_{so} + \cdots 
\end{eqnarray}
for $|\varepsilon_{n}| \ll k_{B}T_{K}$, with $T_{K}$ the 
Kondo temperature discussed in Introduction and defined below.

When polarizations of conduction electrons are ignored, 
it follows according to Yoshimori's\cite{Yoshimori} that 
\begin{equation}\label{EqYoshiPhiM}
\tilde{\phi}_\gamma =
\frac{\tilde{\phi}_c}{2(2l+1)}
+\frac{3\tilde{\phi}_s}{2(2l+3)}
+ \frac{4l(l+1)\tilde{\phi}_o}{(2l+1)(2l+3)} ,
\end{equation}
%
%
and static susceptibilities at $T=0$~K of the Anderson model are
simply given by
\begin{equation}\label{EqA_SusC}
\tilde{\chi}_{c}(0)  \equiv
\frac{\partial \phantom{\Delta\mu}}{\partial \Delta\mu} 
\sum_{m\sigma}\langle n_{m\sigma} \rangle
= \tilde{\chi}_{c}^* (0) ,
\end{equation}
\begin{eqnarray}\label{EqA_SusS}
\tilde{\chi}_{s}(0) &\equiv& 
\frac{\partial \phantom{H_{s}}}{\partial H_{s}} 
\sum_{m\sigma} \frac{1}{2}\sigma g_s\mu_B \langle  n_{m\sigma}
\rangle =\frac{1}{4}g_s^2\mu_B^2 \hspace{1pt}
\tilde{\chi}_{s}^* (0),
\end{eqnarray}
\begin{eqnarray}\label{EqA_SusL}
\tilde{\chi}_o(0) &\equiv& 
\frac{\partial \phantom{H_o}}{\partial H_o} 
\sum_{m\sigma} m g_o\mu_B\langle n_{m\sigma} \rangle
\nonumber \\ &=&
\frac{1}{3}l(l+1) g_o^2\mu_B^2
\hspace{1pt}\tilde{\chi}_o^*(0),
\end{eqnarray}
and
\begin{eqnarray}\label{EqA_SusSL}
\tilde{\chi}_{so}(0) &\equiv& 
\frac{\partial \phantom{H_{so}}}{\partial H_{so}} 
\!\sum_{m\sigma} \frac{1}{2}(m\sigma) g_{so}\mu_B \langle
n_{m\sigma} \rangle
\nonumber \\ &=& 
\frac{1}{4} \cdot
\frac{1}{3}l(l+1) g_{so}^2\mu_B^2 
\hspace{1pt}\tilde{\chi}_{so}^*(0),
\end{eqnarray}
with
\begin{equation}
\tilde{\chi}_{\alpha}^* (0) =
2(2l+1) \tilde{\phi}_{\alpha} \rho(0) ,
\end{equation}
for charge $(\alpha=c)$, spin $(\alpha=s)$, band or orbital
$(\alpha=o)$, and spin-orbital combined  $(\alpha=so)$ channels. 
Here, $\langle \cdots \rangle$ stands for the thermal
average, and
\begin{eqnarray}
\rho(\varepsilon) &=&  
\frac{1}{2l+1} \left(-\frac{1}{\pi}\right)\mbox{Im} \left[ 
\mbox{Tr} \ \tilde{\cal G}_{\sigma}(\varepsilon+i0) \right] 
%
\end{eqnarray}
is the density of states per spin and band.

So far, there is no restriction for $U$, $U^\prime$, or $J$. 
When $U=U^\prime$ and $J=0$, 
$\tilde{\phi}_{s} =\tilde{\phi}_o=\tilde{\phi}_{so}$. It is
reasonable to assume that $U \gg J$, $U^\prime \gg J$ and $U
\agt U^\prime$ so that 
$\tilde{\phi}_{s} \simeq \tilde{\phi}_o \simeq
\tilde{\phi}_{so}$. Only a single energy scale, the Kondo
temperature, appears in low-energy phenomena of the single-site 
problem. In this paper, it is defined by
\begin{equation}\label{EqDefTK}
k_{B}T_{K} = \frac{2l+1}{\tilde{\chi}_{s}^* (0)}
= 2\tilde{\phi}_{s} \rho (0) .  
\end{equation}
The specific heat at low temperatures,
$T \ll T_{K}$,  is given by
$\tilde{C} = \tilde{\gamma} T + \cdots$, with
\begin{equation}
\tilde{\gamma} = \frac{2(2l+1)}{3}
\tilde{\phi}_\gamma \pi^{2} k_{B}^{2} \hspace{1pt} \rho(0).
\end{equation}
The Kondo temperature, $k_{B}T_{K}$,
is as large as the effective Fermi energy of quasiparticles
in the lattice system.

In this paper, the  Kondo temperature is  treated
as a phenomenological parameter
instead of determining and solving 
the Anderson model selfconsistently.

Define a two-point polarization 
function of the lattice system, which is
a $2(2l+1) \times 2(2l+1)$ matrix,  by
\begin{eqnarray}
\bigl[\mbox{\boldmath $\Pi$}(i \omega_l,{\bf q})
\bigr]_{m\sigma,m^\prime\sigma^\prime}
&=&
\frac{1}{N} \sum_{ij}\!\int_{0}^{1/k_{B}T} \hspace{-10pt} d\tau 
\ e^{i \omega_l\tau +i{\bf q}\cdot({\bf R}_{i}-{\bf R}_{j})} 
\nonumber \\ && \times \langle 
n_{mi\sigma}(\tau)n_{m^\prime j\sigma^\prime}(0)\rangle .
\end{eqnarray}
When its irreducible part is denoted  by
$\mbox{\boldmath $\pi$}(i \omega_l,{\bf q})$, 
\begin{eqnarray}
\mbox{\boldmath $\Pi$}
(i \omega_l,{\bf q}) &=&
 \mbox{\boldmath $\pi$}(i \omega_l,{\bf q}) 
\left[1 + {\cal U}
\mbox{\boldmath$\pi$}(i \omega_l,{\bf q}) \right]^{-1} .
\end{eqnarray}
According to Eq.~(\ref{EqIntB}), ${\cal U}$ is defined by 
\begin{equation}\label{EqU3}
\bigl[{\cal U}\bigr]_{m\sigma,m^\prime\sigma^\prime}=
\left\{\begin{array}{cc}
0, &\ m=m^\prime \mbox{~and~} \sigma=\sigma^\prime
\vspace{0.1cm}\\
U^\prime-J, &\ m\ne m^\prime \mbox{~and~} \sigma=\sigma^\prime
\vspace{0.1cm}\\
U, &\ m=m^\prime \mbox{~and~} \sigma\ne\sigma^\prime
\vspace{0.1cm}\\
U^\prime, &\ m\ne m^\prime \mbox{~and~} \sigma\ne\sigma^\prime
\end{array}\right. .
\end{equation}
The irreducible part is divided into a single-site term 
$\tilde{\mbox{\boldmath$\pi$}}(i \omega_l)$ and a multisite
term 
$\Delta\mbox{\boldmath$\pi$}(i \omega_l,{\bf q})$ so that
\begin{equation}
\mbox{\boldmath$\pi$} (i \omega_l,{\bf q}) =
\tilde{\mbox{\boldmath$\pi$}}(i \omega_l) +
\Delta\mbox{\boldmath$\pi$}(i \omega_l,{\bf q}).
\end{equation}
The reducible polarization
function of the mapped Anderson model
is given by
\begin{eqnarray}
\tilde{\mbox{\boldmath$\Pi$}}(i \omega_l) &=&
\tilde{\mbox{\boldmath$\pi$}}(i \omega_l)
\bigl[ 1 + {\cal U}\tilde{\mbox{\boldmath$\pi$}} 
(i \omega_l)\bigr]^{-1} .
\end{eqnarray}
Define the following $2(2l+1) \times 2(2l+1)$ matrix:
\begin{equation}
K^{(\alpha)}_{m\sigma,m^\prime\sigma^\prime}
=\left\{\begin{array}{cl}
\displaystyle \frac{1}{2(2l+1)} \ , & \alpha=c  
\vspace{0.2cm} \\
\displaystyle \frac{1}{2(2l+1)}
\sigma\sigma^\prime\ , & \alpha=s
\vspace{0.2cm}\\ 
\displaystyle \frac{1}{2(2l+1)}
\frac{3mm^\prime}{l(l+1)}\ , &
\alpha=o
\vspace{0.2cm}\\ 
\displaystyle \frac{1}{2(2l+1)}
\frac{3mm^\prime\sigma\sigma^\prime}
{l(l+1)}\ , & \alpha=so
\end{array}\right. .
\end{equation}
Because our system has the particular symmetry such as
Eqs.~(\ref{EqSym}), (\ref{EqSymU}) and (\ref{EqU3}), 
susceptibilities of the four channels are simply given by 
\begin{eqnarray}\label{EqSusL1*}
\chi_\alpha^* (i \omega_l,{\bf q}) &\equiv&
2(2l+1) \!\sum_{m\sigma} \!\sum_{m^\prime\!\sigma^\prime} \!
K^{(\alpha)}_{m\sigma,m^\prime\!\sigma^\prime}
\bigl[\Pi(i \omega_l,{\bf q})
\bigr]_{m\sigma,m^\prime\!\sigma^\prime}
\nonumber \\ &=& 
2(2l+1)\frac{\pi_\alpha(i\omega_l,{\bf q})}
{1- U_\alpha\pi_\alpha(i\omega_l,{\bf q})} 
\end{eqnarray}
and
\begin{eqnarray}\label{EqSusL2*}
\tilde{\chi}_\alpha^* (i \omega_l) &\equiv& 
2(2l+1)\sum_{m\sigma} \sum_{m^\prime\sigma^\prime}
K^{(\alpha)}_{m\sigma,m^\prime\sigma^\prime}
\left[\tilde{\mbox{\boldmath$\Pi$}}(i \omega_l)
\right]_{m\sigma,m^\prime\sigma^\prime}
\nonumber \\ &=&
2(2l+1) \frac{\tilde{\pi}_\alpha(i\omega_l)}
{1- U_\alpha\tilde{\pi}_\alpha(i\omega_l)}, 
\end{eqnarray}
with
\begin{eqnarray}
U_{\alpha} &=& - 
\sum_{m\sigma}\!\sum_{m^\prime\sigma^\prime}
K^{(\alpha)}_{m\sigma,m^\prime\sigma^\prime}
\big[\hspace{1pt} 
{\cal U}\bigr]_{m\sigma,m^\prime\sigma^\prime}
\nonumber \phantom{\Bigg]}\\ &=&
\left\{\begin{array}{cl}
-U-2l \hspace{1pt} (2U^\prime-J), & \alpha=c
\vspace{0.2cm}\\
U + 2l J, & \alpha=s
\vspace{0.2cm}\\
-U+2U^\prime- J, & \alpha=o
\vspace{0.2cm}\\
U - J, & \alpha=so
\end{array}\right. .
\end{eqnarray}
Here, $\pi_{\alpha}(i\omega_l,{\bf q})$ and
$\tilde{\pi}_{\alpha}(i\omega_l)$ are irreducible polarization
functions defined by 
\begin{equation}
\tilde{\pi}_{\alpha}(i\omega_l)  = 
\sum_{m\sigma}\!\sum_{m^\prime\sigma^\prime}
K^{(\alpha)}_{m\sigma,m^\prime\sigma^\prime}
\Bigl[\tilde{\mbox{\boldmath$\pi$}}(i \omega_l) 
\Bigr]_{m\sigma,m^\prime\sigma^\prime},
\end{equation}
\begin{equation}
\Delta\hspace{-1pt}\pi_{\alpha}(i\omega_l,{\bf q}) =
\hspace{-2pt}\sum_{m\sigma}
\hspace{-2pt}\sum_{m^\prime\!\sigma^\prime}
\hspace{-2pt}K^{(\alpha)}_{m\sigma,m^\prime\!\sigma^\prime}
\hspace{-1pt}\Bigl[\hspace{-1pt}
\Delta\hspace{-1pt}\mbox{\boldmath$\pi$}
(i\omega_l,{\bf q}) 
\hspace{-2pt}\Bigr]_{m\sigma,m^\prime\!\sigma^\prime}, 
\hspace{-2pt}
\end{equation}
and
\begin{equation}
\pi_{\alpha}(i\omega_l,{\bf q}) =
\tilde{\pi}_{\alpha}(i\omega_l)
+ \Delta\pi_{\alpha}(i\omega_l,{\bf q}) .
\end{equation}
Note that $\tilde{\chi}_\alpha^* (i \omega_l)$ is nothing
but the susceptibility of the mapped Anderson model.

It follows from Eqs.~(\ref{EqSusL1*}) and (\ref{EqSusL2*})
 that
\begin{equation}\label{EqSusKond}
\chi_{\alpha}^* (i \omega_l,{\bf q}) =
\frac{\tilde{\chi}_{\alpha}^* (i \omega_l) }
{1- \frac{1}{4}I_{\alpha}(i \omega_l,{\bf q})
\tilde{\chi}_{\alpha}^* (i \omega_l) } ,
\end{equation}
with
\begin{equation}\label{EqExch0}
I_{\alpha}(i \omega_l,{\bf q}) =  \frac{2}{2l+1} \cdot
\frac{\Delta \pi_{\alpha}(i \omega_l,{\bf q})}
{\tilde{\pi}_{\alpha}(i \omega_l)
\pi_{\alpha}(i \omega_l,{\bf q})} .
\end{equation}
In this paper, the strong coupling regime is defined by
\begin{equation}\label{EqStrongC}
U_{s} \gg k_{B}T_{K}
\end{equation}
instead of Eq.~(\ref{EqStrongU}). When either of the flat-band
and band-edge conditions is satisfied, Eq.~(\ref{EqStrongC})
is satisfied. Eq.~(\ref{EqStrongC}) leads to a set of  
$U_{\alpha} \tilde{\chi}_{\alpha}^* (0) \gg 1$ and
$U_{\alpha} \chi_{\alpha}^* (0,{\bf q}) \gg 1$ except for
$\alpha=c$. Then, as long as $|\omega_l| \alt k_{B}T_{K}$, 
\begin{equation}\label{EqStrongU1}
U_\alpha \tilde{\pi}_\alpha (i\omega_l) =
1 + O \bigl(1/U_\alpha \tilde{\chi}_\alpha^* (0) \bigr)
\end{equation}
and
%
$U_\alpha \pi_\alpha (i\omega_l, {\bf q}) =
1 + O \bigl(1/U_\alpha \chi_\alpha^* (0, {\bf q}) \bigr)$
%
so that
\begin{equation}\label{EqExch}
I_{\alpha}(i \omega_l,{\bf q}) =  
\frac{2}{2l+1}U_{\alpha}^{2}
\Delta \pi_{\alpha}(i \omega_l,{\bf q}) .
\end{equation}
The susceptibility given by Eq.~(\ref{EqSusKond}) is
consistent with a physical picture for Kondo lattices, where
local fluctuations at different sites interact with each other
by intersite exchange interactions. Then, we call 
$I_{\alpha}(i \omega_l,{\bf q})$ given by Eq.~(\ref{EqExch0}) 
or (\ref{EqExch})  an intersite exchange interaction for  the
spin, orbital, or combined channel. A perturbative treatment
of $I_{\alpha}(i \omega_l,{\bf q})$ is nothing but the $1/d$
expansion; $I_{\alpha}(i \omega_l,{\bf q})$ is of higher order
in $1/d$ for almost all {\bf q}'s, and it is  leading order in
$1/d$ for particular {\bf q}'s such as ${\bf q}=0$, nesting
wave vectors, and so on.

\begin{figure} 
\vspace{0.2cm}~\\
\centerline{\BoxedEPSF{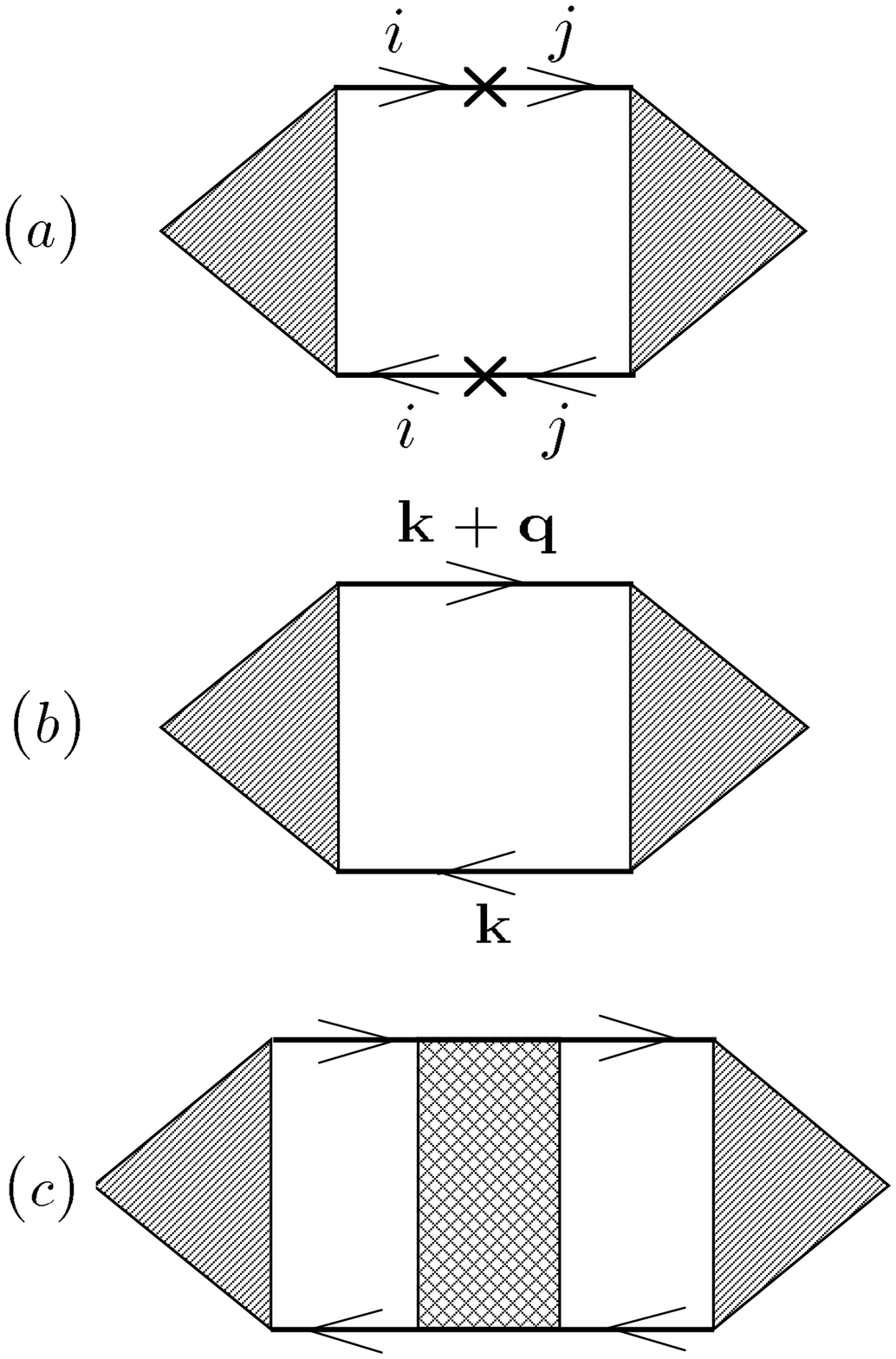 scaled 330}}
\vspace{0.2cm}~\\
\caption[1]{ 
Examples of two-line diagrams for 
$\Delta \pi_\alpha (i\omega_l, {\bf q})$.   Diagram 
$(a)$, which is in the site representation,  gives the
superexchange interaction between the $i$th and
$j$th sites. Diagram $(b)$, which is in the wave-number
representation, gives the novel exchange interaction, which
arises from the virtual exchange of pair excitations
of quasi-particles with wave numbers ${\bf k}+{\bf q}$ and 
${\bf k}$.  Two-line diagrams higher order in intersite effects
such as $(c)$ can be still of leading order in $1/d$, but they
are ignored in this paper. A solid line stands for the
single-particle Green function,  
$\tilde{G}_{m\sigma} (i\varepsilon_n)$ given by
Eq.~(\ref{EqGst}) in $(a)$ or the intersite part of  
${\cal G}_{\sigma} (i\varepsilon_{n},{\bf k}+{\bf q})$ or
${\cal G}_{\sigma} (i\varepsilon_{n},{\bf k})$ given by 
Eqs.~(\ref{EqGrMM}) and (\ref{EqGrMM1}) in $(b)$. A cross in
$(a)$ stands for $t_{mi,m^\prime j}$ given by
Eq.~(\ref{EqTmm}), a shaded triangle for the single-site
irreducible three-point vertex function given by
Eq.~(\ref{EqLambda1}) in $(a)$ or Eq.~(\ref{EqLambda2}) in
$(b)$, and a shaded rectangle in $(c)$ for the single-site
irreducible four-point vertex function.  }
\label{FigTwo-Line}
\end{figure}

In general,  an exchange interaction arises from the virtual
exchange of bosons or bosonic excitations. The exchange
interaction given by Eq.~(\ref{EqExch}) is divided into four
terms in such a way that
\begin{eqnarray}\label{EqIA}
I_\alpha(i\omega_l,{\bf q}) &=&
I_\alpha^{(s)}(i\omega_l,{\bf q})
+I_\alpha^{(Q)}(i\omega_l,{\bf q}) 
\nonumber \\ && \quad 
+ \Delta I_\alpha^{(sQ)} (i\omega_l,{\bf q})
- 4\Lambda_\alpha(i\omega_l,{\bf q})  .
\end{eqnarray}
Only contributions from the so called two-line diagrams can be
of leading order in $1/d$. Three examples of two-line diagrams
are shown in Fig.~\ref{FigTwo-Line}.  The first term,
$I_{\alpha}^{(s)}(i\omega_l,{\bf q})$, corresponds to
Fig.~\ref{FigTwo-Line}$(a)$, and is due to the virtual exchange
of pair excitations of electrons or holes across the
Mott-Hubbard gap, whose excitation energies are as large as
$U$ or $U^\prime\pm J$ and  much larger than $k_{B}T_{K}$. This
is a generalized superexchange interaction. The second term,
$I_{\alpha}^{(Q)}(i\omega_l,{\bf q})$, corresponds to
Fig.~\ref{FigTwo-Line}$(b)$, and is due to the virtual exchange
of pair excitations of quasiparticles, whose excitation
energies are $O(k_{B}T_{K})$. This is the novel exchange
interaction discussed in Introduction. The third term, $\Delta
I_\alpha^{(sQ)} (i\omega_l,{\bf q})$, is the contribution from
two-line diagrams higher order in intersite effects such as
Fig.~\ref{FigTwo-Line}$(c)$. The last term,
$-4\Lambda_\alpha (i\omega_l,{\bf q})$, includes all the other
contributions, which are definitely higher order in $1/d$. Its
lowest-order Feynman diagrams correspond to those of the
mode-mode coupling terms considered in the so called
selfconsistent renormalization (SCR) theory of spin
fluctuations.\cite{MK,Moriya}

The superexchange interaction, 
$I_{\alpha}^{(s)}(i\omega_l,{\bf q})$, is considered in
Sec.~\ref{SecExchSup}, and the novel exchange interaction, 
$I_{\alpha}^{(Q)}(i\omega_l,{\bf q})$, is considered in
Sec.~\ref{SecExchNov}. On the other hand, 
$\Lambda_\alpha(i\omega_l,{\bf q})$ as well as 
$\Delta I_\alpha^{(sQ)}(i\omega_l,{\bf q})$ are ignored in
this paper. Because $\Lambda_\alpha (i\omega_l,{\bf q})$ is of
higher order in $1/d$ for any ${\bf q}$, it can be ignored at
least for large enough $d$.

\subsection{Superexchange interaction}
\label{SecExchSup}

There is no restriction for carrier densities so far. In the
following part of this paper, our study is restricted to low
densities of carriers\cite{ComValence}
\begin{equation}\label{EqLowDensity}
\langle n_{m\sigma} \rangle \ll 1, \quad 
n = \sum_{m\sigma} \langle n_{m\sigma} \rangle < 1,
\end{equation}
so that multiply occupied configurations at a single site can
be ignored in the ground state and low lying excited states.
The single-particle Green function is approximately given
by $\bigl[\tilde{\cal G}_\sigma(i\varepsilon_n)
\bigr]_{mm^\prime}=\delta_{mm^\prime}
\tilde{G}_{m\sigma}(i\varepsilon_{n})$, with
\begin{eqnarray}\label{EqGst}
\tilde{G}_{m\sigma}(i\varepsilon_{n}) &=&
\frac{ \langle n_{m\sigma} \rangle + n_e}
{i\varepsilon_{n}+\mu - \epsilon_{a} + 
i\Gamma\mbox{sgn}(\varepsilon_{n})}
\nonumber \\ && 
+\frac{\langle n_{m-\sigma} \rangle}
{i\varepsilon_{n}+\mu- \epsilon_{a} - U + 
i \Gamma\mbox{sgn}(\varepsilon_{n})}
\nonumber \\ &&  
+ \! \sum_{m^{\prime}\ne m} \Biggl[ \! \frac{ 
\langle n_{m^{\prime}\sigma} \rangle 
+\frac{1}{2} \langle n_{m^{\prime}-\sigma} \rangle} 
{i\varepsilon_{n}+\mu - \epsilon_{a} - U^\prime + J 
+  i \Gamma\mbox{sgn}(\varepsilon_{n})}
\nonumber \\ && 
\hspace*{0.7cm} + \frac{\frac{1}{2}  
\langle n_{m^{\prime}-\sigma} \rangle}
{i\varepsilon_{n}+\mu - \epsilon_{a}- U^\prime-J + 
i \Gamma\mbox{sgn}(\varepsilon_{n})} \Biggr]
\nonumber \\ &&  + \cdots\cdots . \phantom{\Bigr]}
\end{eqnarray}
Here, $\pi\Gamma \simeq 1/\rho_0(0)$ and
\begin{equation}
n_e \simeq 1 - \sum_{m^\prime\sigma^\prime}
\langle n_{m^\prime\sigma^\prime}\rangle
\end{equation}
is the density of empty sites.
This is a straightforward extension of Eq.~(\ref{EqHubTh}) or
Hubbard's theory.\cite{Hubbard}

According to the Ward identity,\cite{Ward}  
the reducible single-site three-point vertex function
is given by 
\begin{eqnarray}\label{EqWard}
\left[\hspace{-1pt}\tilde{\mbox{\boldmath$\Lambda$}}
(i\varepsilon_{n},i\varepsilon_{n};0)
\hspace{-1pt}\right]_{m\sigma,m^\prime\sigma^\prime} 
\hspace{-3pt} &=&
\frac{\partial \tilde{G}_{m\sigma}^{-1}(i\varepsilon_{n})}
{\partial \Delta E_{m^\prime\sigma^\prime}} .
%
\end{eqnarray}
When only the dependence of $\langle n_{m\sigma}\rangle$ on
$\Delta E_{m^\prime\sigma^\prime}$ is considered in
Eq.~(\ref{EqGst}), the irreducible single-site three-point
vertex function is calculated in such a way that
\begin{eqnarray}\label{EqLambda1}
\tilde{\lambda}_{\alpha}
(i\varepsilon_{n},i\varepsilon_{n};0) &=&
\sum_{m\sigma}\sum_{m^\prime\sigma^\prime}
K^{(\alpha)}_{m\sigma,m^\prime\sigma^\prime}
\nonumber \\ && \times
\left[\tilde{\mbox{\boldmath$\Lambda$}}
(i\varepsilon_{n},i\varepsilon_{n};0) [1 - {\cal U}
\tilde{\mbox{\boldmath$\pi$}}(0)]
\right]_{m\sigma,m^\prime\sigma^\prime}
\nonumber \\ &=&
- \frac{\tilde{\pi}_{\alpha}(0)}
{\tilde{G}_{m\sigma}^{2}(i\varepsilon_{n})}
\sum_{r=1}^{4} \frac{A_{\alpha;r}}
{i\varepsilon_{n} -z_{r} + 
i \Gamma \mbox{sgn}(\varepsilon_{n})} ,
\nonumber \\
\end{eqnarray}
with
\begin{equation}
\left(\begin{array}{c}
z_{1} \vspace{0.1cm}\\  
z_{2} \vspace{0.1cm}\\  
z_{3} \vspace{0.1cm}\\  z_{4}
\end{array}\right)
= 
\left(\begin{array}{c}
\epsilon_{a}-\mu \vspace{0.1cm}\\  
\epsilon_{a}+U-\mu \vspace{0.1cm}\\  
\epsilon_{a}+U^\prime-J-\mu \vspace{0.1cm}\\  
\epsilon_{a}+U^\prime+J-\mu
\end{array}\right) 
\end{equation}
and
\begin{equation}
\left( \begin{array}{ccc}
 A_{s;1} &  A_{o;1} &  A_{so:1} 
\vspace{0.1cm}\\ 
 A_{s;2} & A_{o;2} & A_{so:2} 
\vspace{0.1cm}\\  
 A_{s;3} & A_{o;3} & A_{so:3} 
\vspace{0.1cm}\\  
 A_{s;4} & A_{o;4} & A_{so:4} 
\end{array}\!\right)
= \left(\!\begin{array}{rrr} 
 1 & 1 & 1 \vspace{0.1cm}\\  
 -1 & 1 & -1 \vspace{0.1cm}\\   
 l  & \ -\frac{3}{2} & \ -\frac{1}{2} 
\vspace{0.1cm}\\  
 -l &  -\frac{1}{2} & \frac{1}{2}
\end{array}\right) ,
\end{equation}
for spin $(\alpha=s)$, orbital $(\alpha=o)$, and combined 
$(\alpha=so)$ channels.

Consider an exchange interaction between the $i$th and $j$th
sites, which is defined by
\begin{equation}
I_{\alpha;ij}^{(s)} (i\omega_l) =
\frac{1}{N}
\sum_{\bf q} I_{\alpha}^{(s)}(i\omega_l,{\bf q}) 
e^{-i{\bf q}\cdot({\bf R}_{i}-{\bf R}_{j})} ,
\end{equation}
%
%
%
with ${\bf R}_{i}$ being the position of the $i$th
unit cell.
Transfer integrals between the $i$th and $j$th sites have also
the following symmetry:\cite{Com1/d}
\begin{eqnarray}\label{EqTmm}
t_{mi,m^\prime j} &=& - \frac{1}{N}
\sum_{\bf k} E_{mm^\prime}({\bf k}) 
e^{i {\bf k} \cdot ({\bf R}_{i}-{\bf R}_{j})} 
\nonumber \\ &\equiv&
\left\{\begin{array}{ll}
 t_{ij}, & m=m^\prime \vspace{0.2cm}\\
 t_{ij}^\prime, & m \ne m^\prime
\end{array} \right. .
\end{eqnarray}
When the two-line diagram shown in Fig.~\ref{FigTwo-Line}$(a)$
is taken into account and Eq.~(\ref{EqStrongU1}) is made use
of, the superexchange interaction
is calculated in such a way that  
\begin{equation}\label{EqSup}
I_{\alpha;ij}^{(s)}(0) =
\frac{2}{2l+1} \cdot \frac{2 }{U} \ T_{ij}^{(\alpha)}
Y_{\alpha;ij} ,
\end{equation}
with
\begin{eqnarray}\label{EqTij}
T_{ij}^{(\alpha)} &=& 
\sum_{mm^\prime\sigma}
K^{(\alpha)}_{m\sigma,m^\prime\sigma}
|t_{mi,m^\prime j}|^{2} 
\nonumber \\ &=&
\left\{ \begin{array}{ll}
|t_{ij}|^2 + 2l |t_{ij}^\prime|^2 , &\alpha = s \vspace{0.2cm}\\
|t_{ij}|^2-|t_{ij}^\prime|^2 , & \alpha = o \mbox{~and~} so 
\end{array} \right. 
\end{eqnarray}
and
\begin{eqnarray}\label{EqYij0}
Y_{\alpha;ij} &=&
\frac{1}{2} U\left(-  k_{B}T\right) \sum_{\varepsilon_{n}}
 U_\alpha^2 \tilde{\lambda}_\alpha^2 
(i\varepsilon_{n}, i\varepsilon_{n}; 0)
\tilde{G}_{m\sigma}^4(i\varepsilon_{n})
\nonumber \\ &=& 
\frac{1}{2} U
\sum_{r=1}^4 \sum_{r^\prime=1}^4 
A_{\alpha; r}A_{\alpha; r^\prime} 
\Xi(z_{r},z_{r^\prime};\Gamma) .
\end{eqnarray}
Here, $\Xi(z,z^\prime;\Gamma)$ is defined in such a way that
\begin{equation}
\Xi(z,z^\prime;\Gamma) \equiv \frac{1}{\pi(z\!-\!z^\prime)} 
\!\!\left[ \mbox{Tan}^{-1}\hspace{-3pt}\left(\!
\frac{z^{\phantom{\prime}}}{\Gamma}
\!\right) \!-\!
 \mbox{Tan}^{-1}\hspace{-3pt}\left(\! \frac{z^\prime}{\Gamma}
\right)
\!\right] 
\end{equation}
for $z \ne z^\prime$, and
$\Xi(z,z) = \lim_{z^\prime \rightarrow z} \Xi(z,z^\prime)$.
Note that $Y_{\alpha;ij}=-1$ for $\Gamma/U\rightarrow +0$ and
$J=0$.
When  the carrier density is so small that
Eq.~(\ref{EqLowDensity})  is satisfied,  the contribution from
excitations within  the lower Hubbard band $(r=r^\prime=1)$ is
nothing but that from the virtual exchange of pair excitations
of quasi-particles. Because it is considered in the next
subsection, it should be subtracted in order to avoid double
counting so that Eq.~(\ref{EqYij0}) is replaced
by\cite{ComSubtract}
\begin{equation}\label{EqYij}
Y_{\alpha;ij} = \frac{1}{2}U \sum_{r\ne r^\prime}
A_{\alpha; r}A_{\alpha; r^\prime}
\Xi (z_{r},z_{r^\prime};\Gamma) .
\end{equation}
In the following part, 
$I_{\alpha;ij}^{(s)}(i\omega_l=0)$ is simply denoted
by $I_{\alpha;ij}^{(s)}$. 
Because its $\omega_l$ dependence is small,
it can be used for $|\omega_l| \ll U$ and $U^\prime-J$.

In the small limit of $\Gamma/U\rightarrow +0$, 
it follows that
\begin{mathletters}\label{EqIij}
\begin{equation}\label{EqIijS}
I_{s;ij}^{(s)} =
-\frac{\displaystyle 4 T_{ij}^{(s)}
}{2l+1} 
\!\left[ \frac{1}{U}-l \left(\!\frac{1}{U^\prime\!-\!J}
\!-\! \frac{1}{U^\prime \!+\! J} \!\right)\!\right]  ,
\end{equation}
\begin{equation}\label{EqIijT}
I_{o;ij}^{(s)} =
-\frac{\displaystyle 4 T_{ij}^{(o)}
}{2l+1} 
\!\left[-\frac{1}{U}+\frac{1}{2}
\!\left(\!\frac{3}{U^\prime\!-\! J}
\!+\! \frac{1}{U^\prime \!+\! J} \!\right)\!\right]  ,
\end{equation}
and
\begin{equation}\label{EqIijST}
I_{so:ij}^{(s)} =
- \frac{\displaystyle 4 T_{ij}^{(so)}
}{2l+1} 
\!\left[ \frac{1}{U} + \frac{1}{2}
\!\left(\!\frac{1}{U^\prime\!-\!J}
\!-\! \frac{1}{U^\prime \!+\! J} \!\right)\!\right]  .
\end{equation}
\end{mathletters}
In deriving these results,
$\epsilon_a -\mu \ll - k_BT$, $\epsilon_a +U-\mu \gg k_BT$ 
and $\epsilon_a +U^\prime -J -\mu \gg k_BT$ are assumed. These
results agree with those calculated for insulating phases  by
the conventional perturbative method.\cite{Roth}

Note that Eq.~(\ref{EqIijS}) can be applied even to $l= 0$. The
superexchange interaction in spin channels can never be
ferromagnetic in the single-band model $(l=0)$.
It can be ferromagnetic in multi-band models
$(l\ge 1/2)$, when $J$ is as large as
\begin{equation}\label{EqJC1}
J > \sqrt{\displaystyle l^2U^{2}+\left(U^\prime\right)^2}- lU .
\end{equation}
It is interesting that as long as 
$J>0$ the superexchange interaction is definitely
ferromagnetic  in the large limit of orbital multiplicity
$(l \rightarrow + \infty)$.

\begin{figure} 
\centerline{\BoxedEPSF{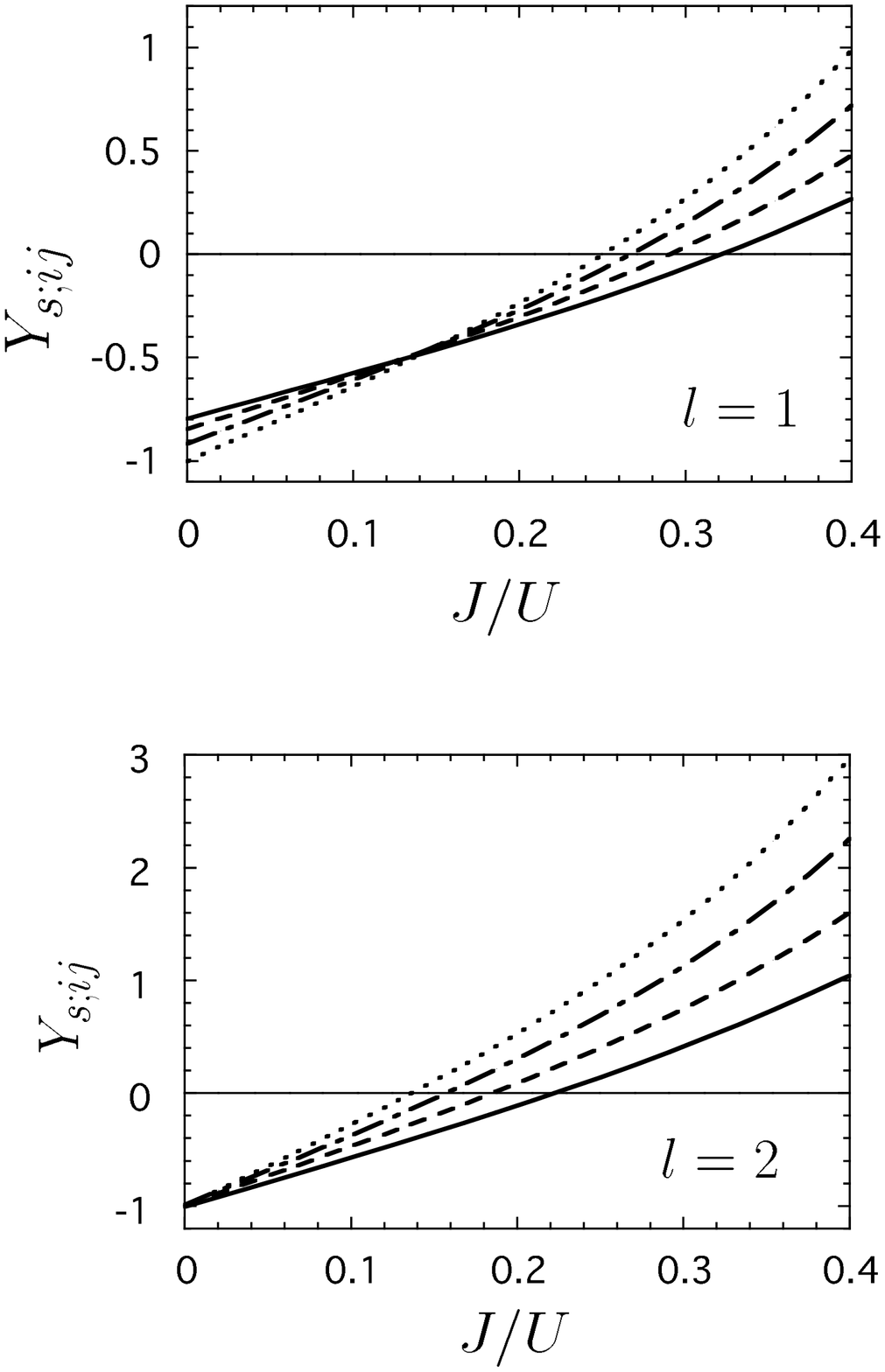 scaled 350}}
~\\
\caption[2]{ 
$Y_{s;ij}$ defined by Eq.~(\ref{EqYij}) as a function of the
Hund coupling,
$J$, for $l=1$ and $l=2$. Here, 
$(\epsilon_a-\mu)/U=-0.1$ and 
$U^\prime/U=3/4$ are assumed. 
A solid line shows $Y_{s;ij}$ for $\pi\Gamma/U=0.3$,
a broken line for $\pi\Gamma/U=0.2$,
 a dotted-broken line for $\pi\Gamma/U=0.1$, and
 a dotted line for $\pi\Gamma/U=0$.
}
\label{FigJ(L=1*2)}
\end{figure}

\begin{figure} 
\centerline{\BoxedEPSF{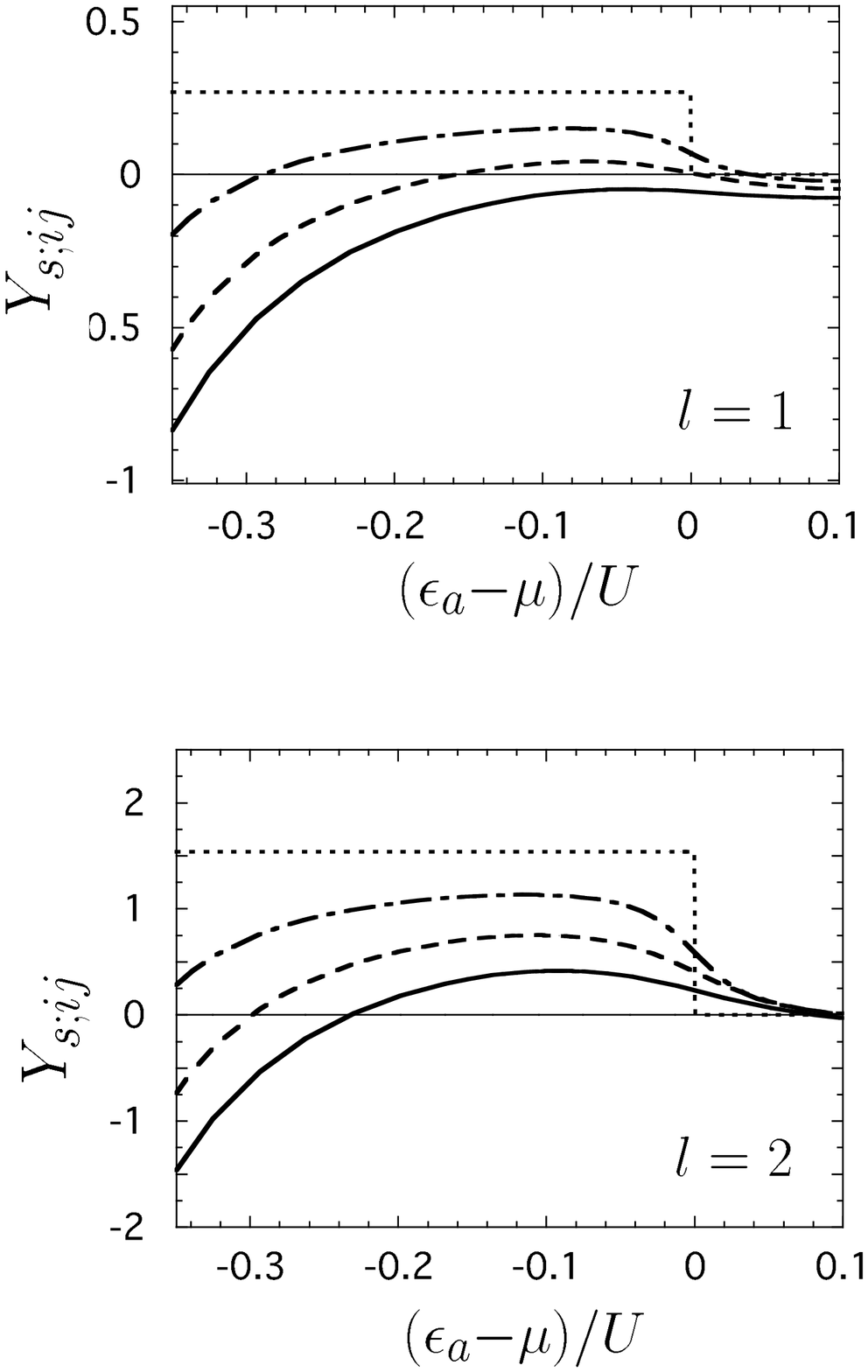 scaled 360}}
~\\
\caption[3]{ 
$Y_{s;ij}$ defined by Eq.~(\ref{EqYij}) as a function of the
depth, $\epsilon_a-\mu$, for $l=1$ and $l=2$. Here, 
$J/U=0.3$ and $U^\prime/U=3/4$ are assumed. A solid line shows
$Y_{s;ij}$ for $\pi\Gamma/U=0.3$, a broken line for
$\pi\Gamma/U=0.2$, a dotted-broken line for $\pi\Gamma/U=0.1$,
and a dotted line for $\pi\Gamma/U=0$.
}
\label{FigD(L=1*2)}
\end{figure}

Non-zero $\Gamma$ makes the superexchange interaction reduced. 
Fig.~\ref{FigJ(L=1*2)} shows $Y_{s;ij}$ as a function of $J$
for various $\Gamma$'s, and  Fig.~\ref{FigD(L=1*2)} shows
$Y_{s;ij}$ as a function of $\epsilon_a-\mu$ for the same
$\Gamma$'s.  Positive and negative $Y_{s;ij}$'s mean that the
superexchange interaction is ferromagnetic and
antiferromagnetic, respectively. When $J$ and $l$ are large
enough, the superexchange interaction is ferromagnetic even
for non-zero $\Gamma$.

The signs of $I_{o;ij}^{(s)}$ and $I_{so:ij}^{(s)}$ depend on
$|t_{ij}|^{2}-|t_{ij}^\prime|^2$.
When $|t_{ij}| \simeq |t_{ij}^\prime|$,  both of
$I_{o;ij}^{(s)}$ and  $I_{so:ij}^{(s)}$ are small.
Interesting is the case where either of $|t_{ij}|^{2}$
and $|t_{ij}^\prime|^2$ is much larger than the other, because 
it is possible that an instability of the orbital or combined
channel occurs at low enough channels. In general, 
$|I_{o;ij}^{(s)}| > |I_{so;ij}^{(s)}|$.

\subsection{Novel exchange interaction}
\label{SecExchNov}

Selfenergy corrections from multi-site effects are of
higher order in $1/d$ except for those due to Weiss' mean
fields, which are not present in Landau's normal Fermi-liquid
phases. When only the single-site selfenergy
given by Eq.~(\ref{EqSelf}) is taken into account, the
single-particle Green function is approximately given by
\begin{equation}
\left[ {\cal G}_{\sigma}^{-1} (i\varepsilon_{n},{\bf k}) 
\right]_{mm^\prime} = \tilde{\phi}_\gamma
\left[ i \varepsilon_{n} \delta_{mm^\prime} - 
e_{mm^\prime}({\bf k}) \right],
\end{equation}
with
\begin{equation}
e_{mm^\prime}({\bf k}) =\frac{1}{\tilde{\phi}_\gamma}
\left[ E_{mm^\prime}({\bf k}) +
\left(\tilde{\Sigma}_0(0) - \mu \right) \delta_{mm^\prime}
\right] ,
\end{equation}
%
for $|\varepsilon_{n}| \alt k_{B}T_{K}$.
The dispersion relation of quasiparticles, 
$\xi_m({\bf k})$, is determined from 
%
$\mbox{det}\!\left[ {\cal G}_{\sigma}^{-1} 
(\xi_m({\bf k}),{\bf k})\right]  =0 $,
%
so that
\begin{equation}
\xi_m({\bf k}) = \frac{1}{\tilde{\phi}_\gamma}\left[ 
\tilde{\Sigma}_0(0) + E_m^*({\bf k})  -\mu \right] ,
\end{equation}
with $E_m^* ({\bf k})$ given by Eq.~(\ref{EqDisper}).
Note that $\xi_m({\bf k})=\xi_{-l}({\bf k})$ for $m$ such as
$-l \le m \le (l-1)$ and only 
$\xi_l({\bf k})$ can be different from them.
The $(mm^\prime)$th component of the Green function, 
is calculated in such a way that
\begin{equation}\label{EqGrMM}
\left[{\cal G}_{\sigma} (i\varepsilon_{n},{\bf k})
\right]_{mm}  \!=
\frac{1}{(2l\!+\!1)\tilde{\phi}_\gamma}
\!\left[\!\frac{2l}{i\varepsilon_{n}\!\!-\xi_{-l}({\bf k})}
\!+\!
\frac{1}{i\varepsilon_{n}\!-\xi_l({\bf k})}\!\right]
\end{equation}
for intraband components, and
\begin{equation}\label{EqGrMM1}
\left[{\cal G}_{\sigma} (i\varepsilon_{n},{\bf k})
\right]_{mm^\prime} \! =
\frac{(-1)^{2l+1}}{(2l\!+\!1)\tilde{\phi}_\gamma}
\!\left[\!\frac{1}{i\varepsilon_{n}\!\!-\xi_{-l}({\bf k})}
\!-\!
\frac{1}{i\varepsilon_{n}\!-\xi_l({\bf k})}\!\right]
\end{equation}
for interband components $(m\ne m^\prime)$.

According to the Ward identity,\cite{Ward}
\begin{eqnarray}\label{EqLambda2}
 \tilde{\lambda}_{\alpha}(0,0;0)
&=&  [1- U_{\alpha} \tilde{\pi}_{\alpha}(0)]
\tilde{\phi}_{\alpha}
\nonumber \\ &=& 
\frac{2(2l \!+\! 1) 
\tilde{\phi}_{\alpha}}{U_{\alpha}\tilde{\chi}^*_{\alpha}(0)}
\!\left[1 + O \!\left( \!\frac{1}
{U_{\alpha}\tilde{\chi}_{\alpha}^*(0)}
\! \right) \!\right] .
\end{eqnarray}
When this is approximately used for non-zero energies such as
$|\varepsilon_{n}|\alt k_{B}T_{k}$ and $|\omega_l|\alt
k_{B}T_{k}$ and the two-line diagram shown in
Fig.~\ref{FigTwo-Line}$(b)$ is taken into account, the exchange
interaction due to the virtual exchange of pair excitations of
quasiparticles is calculated so that 
\begin{eqnarray}\label{EqIQ}
I_{\alpha}^{(Q)}(i\omega_l,{\bf q})  
&=&  \!\left[ U_{\alpha}\tilde{\lambda}_{\alpha}(0,0;0) 
\! \right]^{2} \!\! \frac{1}{\tilde{\phi}_\gamma^2} \!
\left[P_{\alpha}(i\omega_l,{\bf q}) \!-\!
\tilde{P}_{\alpha} (i\omega_l) \right]
\nonumber \\ &=&
 \left(\frac{\tilde{\phi}_\alpha}{\tilde{\phi}_\gamma }
\right)^2 \!
\frac{8(2l+1)}{[\tilde{\chi}^*_\alpha(0)]^2} \!
\left[ P_\alpha (i\omega_l,{\bf q}) -
\tilde{P}_\alpha (i\omega_l) \right] .
\nonumber \\
\end{eqnarray}
Here, 
\begin{eqnarray}\label{EqPolQ} 
P_{\alpha}(i\omega_l,{\bf q}) &=&
-k_BT \sum_{\varepsilon_{n}}
\frac{1}{N} \sum_{\bf k} \sum_{mm^\prime\sigma} 
K^{(\alpha)}_{m\sigma,m^\prime\sigma} \tilde{\phi}_\gamma^2 
\nonumber \\ && \times
\bigl[{\cal G}_{\sigma}
(i\varepsilon_{n} \!\!+\! i\omega_l,{\bf k} \!+\! {\bf q})
\bigr]_{mm^\prime} \!
\bigl[{\cal G}_{\sigma}
(i\varepsilon_{n},{\bf k})\bigr]_{m^\prime m}
\nonumber \\ && \phantom{\Bigr]}
\end{eqnarray}
is a  a polarization function of quasi-particles and 
\begin{equation}\label{EqPolQL} 
\tilde{P}_\alpha(i\omega_l) =
\frac{1}{N} \sum_{\bf q} 
P_{\alpha}(i\omega_l,{\bf q}) 
\end{equation}
is its local part. Because any single-site term is considered
in the  {\it unperturbed} state, the single-site part is
subtracted in Eq.~(\ref{EqIQ}).

For the spin channel $(\alpha=s)$, Eqs.~(\ref{EqPolQ}) and
(\ref{EqPolQL}) become so simple as
\begin{equation}\label{EqIQS} 
P_{s}(i\omega_l,{\bf q}) = 
\frac{1}{(2l\!+\!1)N} \!\sum_{m{\bf k}} 
\frac{ f[\xi_{m}({\bf k} \!+\! {\bf q})] 
 \!-\! f[\xi_{m}({\bf k})]}
{\xi_{m}({\bf k}) \!-\! \xi_{m}({\bf k} \!+\! {\bf q}) 
\!-\! i \omega_l } 
\end{equation}
and
\begin{equation}\label{EqPLocal}
\tilde{P}_s (0) =
\int_{-\infty}^{\infty} \!\!\!\! dx \!\!
\int_{-\infty}^{\infty} \!\!\!\! dy
\frac{f(y)-f(x)}{x-y} \rho^* (x) \rho^* (y) ,
\end{equation}
with
\begin{equation}\label{EqRho**}
\rho^* (\epsilon) = \frac{1}{2l+1}\sum_m 
\delta \bigl(\epsilon- \xi_m (\epsilon) \bigr)  
\end{equation}
the density of quasi-particle states per spin and band.
For orbital $(\alpha=o)$ and combined $(\alpha=so)$ channels,
\begin{eqnarray}\label{EqIQST} 
P_{\alpha}(i\omega_l,{\bf q}) &=& 
- \frac{2l-1}{(2l\!+\!1)N} \!\sum_{\bf k} 
\frac{ f[\xi_l({\bf k} \!+\! {\bf q})] 
 \!-\! f[\xi_l({\bf k})]}
{\xi_l({\bf k}) \!-\! \xi_l({\bf k} \!+\! {\bf q}) 
\!-\! i \omega_l } 
\nonumber \\ && \hspace*{-1.2cm}
+ \frac{8l}{(2l\!+\!1)N} \!\sum_{\bf k} 
\frac{ f[\xi_{-l}({\bf k} \!+\! {\bf q})] 
 \!-\! f[\xi_l({\bf k})]}
{\xi_l({\bf k}) \!-\! \xi_{-l}({\bf k} \!+\! {\bf q}) 
\!-\! i \omega_l } 
\nonumber \\ &&  \hspace*{-1.2cm}
+ \frac{2l(2l-1)}{(2l\!+\!1)N} \!\sum_{\bf k} 
\frac{ f[\xi_{-l}({\bf k} \!+\! {\bf q})] 
 \!-\! f[\xi_{-l}({\bf k})]}
{\xi_{-l}({\bf k}) \!-\! \xi_{-l}({\bf k} \!+\! {\bf q}) 
\!-\! i \omega_l } .
\end{eqnarray}

This exchange interaction includes no intrasite part  so that 
$\sum_{\bf q} I_{\alpha}^{(Q)}(i\omega_l,{\bf q}) =0$.
When $I_{\alpha}^{(Q)}(0,+0) <0$,
$I_{\alpha}^{(Q)}(0,{\bf q})$ takes its maximum value at
non-zero ${\bf q}$. In this paper, it is classified as being
antiferromagnetic  when $I_{\alpha}^{(Q)}(0, +0) <0$ and  as
being ferromagnetic  when $I_{\alpha}^{(Q)}(0, +0) >0$. 
As was studied in the previous paper,\cite{Miyai}
the temperature dependence of 
$I_{\alpha}^{(Q)}(0,{\bf q})$
is responsible for the Curie-Weiss law.
It is easy to see from Eq.~(\ref{EqIQ}) or (\ref{EqIQ0}) 
that  $I_{\alpha}^{(Q)}(0,{\bf q})$
is of the order of $k_BT_K$.

\section{Flat-band and band-edge models}
\label{SecModel}
\subsection{Instability condition } 

Assume $T=0$~K in this section. 
Then, it follows that 
$P_s (0,|{\bf q}|\rightarrow +0) = \rho^* (0)$ and 
\begin{equation}\label{EqIQ0}
\frac{1}{4} I_{s}^{(Q)}(0,+0)  =
\frac{\tilde{\phi}_s}{\tilde{\phi}_\gamma } 
\left[1-
\frac{\tilde{P}_s (0)}{\rho^*(0) } \right]
\frac{1}{\tilde{\chi}^*_s (0)} .
\end{equation}
Note that $\rho^*(0) = \tilde{\phi}_\gamma \rho (0)$
to leading order in $1/d$.
An instability condition  against ferromagnetism 
is simply given by 
$1  - (1/4) I_s (0,0)\tilde{\chi}_{s}(0)< 0 $ or
\begin{equation}\label{EqFerroCond}
1- \frac{\tilde{\phi}_s}{\tilde{\phi}_\gamma} 
\left[1-\frac{\tilde{P}_s (0)}{\rho^*(0)} 
\right] - \Delta X_s < 0 ,
\end{equation}
with
\begin{equation}\label{EqXs}
\Delta X_s = \frac{1}{4}\left[ I_s^{(s)} (0,0) 
- 4\Lambda_s(0,0) \right] \tilde{\chi}_{s}(0).
\end{equation}
When $J$ is strong enough, the superexchange interaction,
$I_s^{(s)} (0,0) $, is ferromagnetic, as is discussed in
Sec.~\ref{SecExchSup}. The mode-mode coupling, $\Lambda_s(0,0)$,
can be ignored in infinite dimensions
$(d \rightarrow+\infty)$.  It is likely that $\Delta X_s>0$ when
$d$, $l$ and $J$  are large enough. When Eq.~(\ref{EqXs}) is
positively large enough, ferromagnetic instability occurs
even when neither of the flat-band and band-edge conditions is
satisfied. The SCR theory\cite{Murata,MK,Moriya} can only be
applied to such ferromagnets in finite dimensions
$(d\ll+\infty)$. Then, it is examined in this
section whether or not ferromagnetic instability is possible
even when $\Delta X_s$ is as small as 
$|\Delta X_s| \ll k_BT_K$ or $\Delta X_s=0$.

The signs of Eq.~(\ref{EqIQ0}) or 
$\bigl[1- \tilde{P}_s (0)/\rho^*(0) \bigr]$ are functionals of
$\rho^* (\epsilon)$. In order to examine the flat-band
and band-edge conditions, we consider the following model of
$\rho^* (\epsilon)$: It is given by the sum of a narrow
triangular one and  a wide square one and the chemical potential
is just at the peak of the triangle in such a way that
\begin{equation}\label{EqRho}
\rho^* (\epsilon) =
 c \rho_n^* (\epsilon) + 
(1-c) \rho_w^* (\epsilon) ,
\end{equation}
where $\rho_n^* (\epsilon)$ and $\rho_w^* (\epsilon)$ are given
by 
\begin{equation}
\rho_n^* (\epsilon) =
\left\{\begin{array}{cl}
\displaystyle 
\!\!\frac{2}{w_n} \!\left(1 \!-\! 
\left|\frac{\epsilon}{ e_{nd}}\right| \right) ,
 & e_{nd} \le \epsilon \le 0 \\
 & \\
\displaystyle 
\!\!\frac{2}{w_n} \!\left(1 \!-\! 
\left|\frac{\epsilon}{ e_{nu}}\right| \right) ,
 & 0 \le \epsilon \le e_{nu}  \\
 & \\
0\hspace{1pt}, &  \epsilon < e_{nd} , \ e_{nu} < \epsilon
\end{array} \right. ,
\end{equation}
with $w_n = e_{nu}-e_{nd}$, and 
\begin{equation}
\rho_w^* (\epsilon) =
\left\{\begin{array}{cl}
\displaystyle 
\frac{1}{w_{w}}\hspace{1pt}, &  
e_{wd}  \le \epsilon \le e_{wu} \\
 & \\ 
0\hspace{1pt}, &  \epsilon<e_{wd} , \   e_{wu} < \epsilon 
\end{array} \right. ,
\end{equation}
with  $w_w=e_{wu}-e_{wd}$. 
For the sake of simplicity, our examination is restricted to a
realistic case, where a flat band lies
within broad bands: $e_{nd}<0<e_{nu}$, $e_{wd}<0<e_{wu}$,
and 
$w_w \gg w_n$ so that 
\begin{equation}
\rho_n^* (0) \gg \rho_w^* (0) .
\end{equation}
It is also assumed that the quasi-particle density of states is 
normalized in such a way that
\[
\int_{-\infty}^{+\infty} d \epsilon 
\rho^* (\epsilon ) = 1 .
\]
Then, the carrier density per site is given by
\begin{eqnarray}
n &=& 2(2l+1) \int_{-\infty}^{+\infty} \!\! d\epsilon 
f(\epsilon ) \rho^* (\epsilon)
\nonumber \\
&=& 2(2l+1)\left[c \frac{|e_{nd}|}{w_{n}} + 
(1- c) \frac{|e_{wd}|}{w_{w}}\right] .
\end{eqnarray}
Here, Luttinger's sum rule or the Fermi-surface sum
rule\cite{Luttinger1,Luttinger2} is made use of.

In Eq.~(\ref{EqRho}), $c$ is the weight of the triangular part.
However, it is never a free parameter in this paper. When
Eq.~(\ref{EqFlat}) is satisfied, our model is nothing but the
flat-band model. Such a flat-band model is simulated by
$\rho^*(\epsilon)$ with $c \simeq 1/(2l+1) $. When interband
transfer terms $E_{l-l}({\bf k})$ are small, on the other
hand, all the $2l+1$ bands are degenerate. Such a model is
simulated by $\rho^*(\epsilon)$ with $c=1$ or $c = 0$.

It follows that  
\begin{eqnarray}
\tilde{P}_s (0) &=&
 c^2 \rho_n^* (0)   A_n \!\!\left(\! 
\frac{|e_{nd}|}{w_n}\right)
+ (1-c)^2 \rho_w^* (0) 
A_w \!\!\left(\!\frac{|e_{wd}|}{w_w}\right)
\nonumber \\ &&  + 
c(1-c) \rho_w^* (0)  \Biggl[ \frac{2e_{nu}}{w_n} 
A_{nw}\!\!\left(\frac{|e_{wd}|}{e_{nu}}\right) +
\nonumber \\ && \hspace*{2.5cm} + 
 \frac{2|e_{nd}|}{w_n} 
A_{nw}\!\!\left(\frac{ e_{wu} }{|e_{nd}|} \right) \Biggr],
\end{eqnarray}
with
\begin{eqnarray}
A_n (x) &=& - \frac{2}{3}
\Biggl\{1 + \frac{x \bigl[ x + 3(1-x)\bigr]}{1-x} \ln x 
\nonumber \\ && \hspace*{12pt}
+ \frac{(1-x) \bigl[(1-x)+3x \bigr]}{x}\ln (1-x) \Biggr\},
\end{eqnarray}
\begin{equation}
A_{w}(x) = - 2\bigl[ x\ln x +(1-x)\ln (1-x) \bigr],
\end{equation}
and
\begin{equation}
A_{nw}(x)
= \ln x + (x+1)^2\ln\left(1 +\frac{1}{x}\right)-x .
\end{equation}

\begin{figure} 
\centerline{\BoxedEPSF{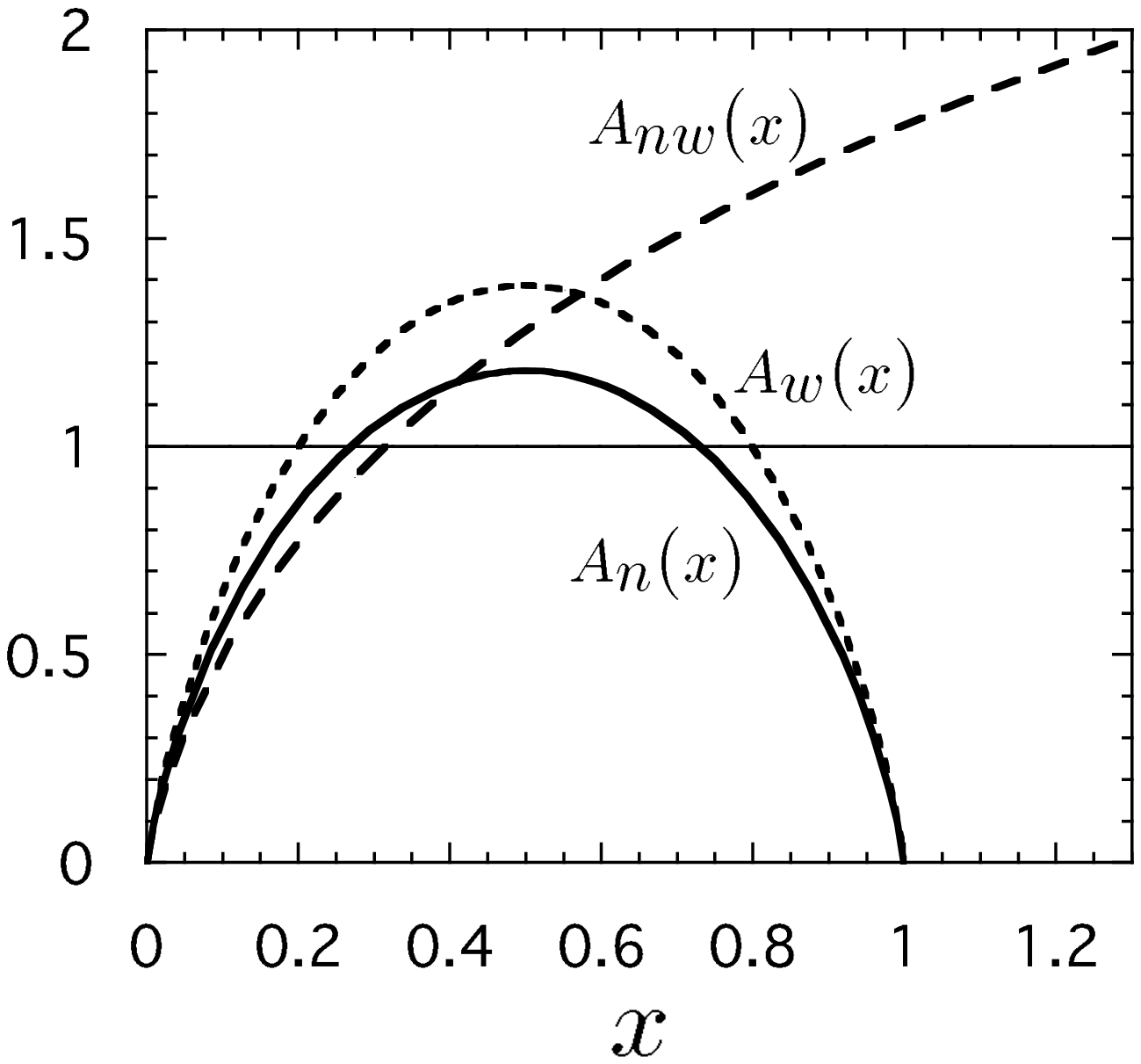 scaled 370}}
\vspace{-0.0cm}~\\
\caption[4]{ 
A solid line shows $A_n(x)$, a dotted line $A_w(x)$,
and a broken line $A_{nw}(x)$.}
\label{Fig-A(x)}
\end{figure}

The instability condition (\ref{EqFerroCond}) is more easily
satisfied for smaller $\tilde{P}_s (0)$.  Fig.~\ref{Fig-A(x)}
shows $A_n(x)$, $A_w(x)$, and $A_{nw}(x)$. Note that 
$A_{n}(0)= A_{n}(1) =0$; $A_{w}(0) = A_{w}(1) =0$;
$A_{nw}(0)=0$ and $A_{nw}(x)$ increases only logarithmically
with increasing $x$ in such a way that 
\begin{equation}
A_{nw}(x) = \ln x + \frac{3}{2} + \frac{1}{3x} -
\frac{1}{12x^2} + \cdots 
\end{equation}
for $x\gg 1$. When the chemical potential is around one of the
band-edges, ferromagnetism is favored. On the other hand,
$A_n(x)$ and $A_w(x)$ take their largest values at $x=1/2$;
$A_n(1/2)=-(2/3)+(8/3)\ln 2=1.1873\cdots$ and 
$A_w(1/2) = 2\ln 2  =1.38629\cdots$. When the chemical
potential is around the band center, ferromagnetism is
disfavored. This is one of the reasons why the band-edge
condition is desired to be satisfied for itinerant-electron
ferromagnetism.

\subsection{Flat-band model}

Consider  the limit of $w_w / w_n\rightarrow +\infty$ or
\begin{equation}\label{EqWideLim}
\rho_w^* (0) / \rho_n^* (0) \rightarrow +0 
\end{equation}
in the case of nonzero $c$ $(c >0)$.  In this limit, 
\begin{equation}\label{EqPsR}
\tilde{P}_s (0) / \rho^*(0)  =
c A_n \left(\! 
|e_{nd}| / w_n \right) .
\end{equation}
Note that $c \simeq 1/(2l+1)$ is smaller than unity. This is
one of the reasons why ferromagnetism is favored in  the
flat-band model. Eq.~(\ref{EqPsR}) vanishes in the limit of 
\begin{equation}\label{EqL-Lim}
l \rightarrow+\infty
\end{equation}
or 
\begin{equation}\label{EqEdgeLim}
|e_{nd}| / w_n  \rightarrow +0.
\end{equation}
Eq.~(\ref{EqEdgeLim}) is nothing but the band-edge condition.
Then, the instability condition (\ref{EqFerroCond})
with $\Delta X_s =0$ becomes quite simple as 
\begin{equation}\label{EqFerroCond1}
1 - \bigl(\tilde{\phi}_s / \tilde{\phi}_\gamma
\bigr) < 0 .
\end{equation}

In the Kondo problem, $\tilde{\phi}_s/\tilde{\phi}_\gamma$
is called the Wilson ratio. Call it the {\it local}\hspace{1pt}
Wilson ratio in this paper. In general, 
it is close to unity in the low density limit of 
$n \rightarrow +0$ and is larger than unity for $n > 1$. 
The carrier density is simply given by
\begin{equation}
n = 2 (2l+1) \left(|e_{wd}|/w_w \right) .
\end{equation}
When it is close to unity, $ n \simeq 1$,
charge fluctuations
are substantially suppressed, so that
$\tilde{\phi}_c \ll \tilde{\phi}_s$ and
$\tilde{\phi}_c \ll \tilde{\phi}_o$, as is examined in
Appendix~\ref{SecApp}.
Then, it follows from Eq.~(\ref{EqYoshiPhiM}) that
\begin{equation}\label{EqYoshiPhiM2}
\tilde{\phi}_\gamma \simeq 
\frac{3\tilde{\phi}_s}{2(2l+3)}
+ \frac{4l(l+1)\tilde{\phi}_o}{(2l+1)(2l+3)}.
\end{equation}
In the case of $U=U^\prime$ and $J=0$,
it follows that  
$\tilde{\phi}_s = \tilde{\phi}_o$, so that
the {\it local}\hspace{1pt} Wilson ratio is as large as 
\begin{equation}\label{EqWR}
\frac{\tilde{\phi}_s}{\tilde{\phi}_\gamma} = 1+
\frac{5l+3}{8l^2 + 11l +3} >1 .
\end{equation}
Here, $\tilde{\phi}_c =0$ is assumed. When $U > U^\prime$ or
$J>0$,  it follows that  $\tilde{\phi}_s > \tilde{\phi}_o$. 
The  {\it local}\hspace{1pt} Wilson ratio is larger than
Eq.~(\ref{EqWR})  so that the instability condition,
Eq.~(\ref{EqFerroCond}) or (\ref{EqFerroCond1}), is satisfied
in such a case. As long as the Hund coupling is nonzero
$(J>0)$, the ferromagnetic instability condition is satisfied 
in the flat-band model with large enough $l$ and $d$.

\subsection{Band-edge model}

Consider the low density limit of $n \rightarrow +0$ in the
case of $c=1$ or $c=0$. According to the above argument,  the
chemical potential should be at the band bottom.   It follows
that $\tilde{P}_s (0) \rightarrow +0$ in the limit of
$n \rightarrow +0$. The instability condition is also given by
Eq.~(\ref{EqFerroCond1}).

In the single-band model, 
$\lim_{n \rightarrow +0}
\bigl(\tilde{\phi}_s/\tilde{\phi}_\gamma\bigr) \simeq  1 $ but 
$\lim_{n \rightarrow +0}
\bigl(\tilde{\phi}_s/\tilde{\phi}_\gamma\bigr) <  1 $ as is
shown in Appendix~\ref{SecApp}.
The superexchange interaction is antiferromagnetic in general,
and it vanishes in the limit of
$U/k_BT_k \rightarrow +\infty$. The single-band model with
$n \rightarrow 0$ and $U/k_BT_K \rightarrow +\infty$ is close
to a quantum critical point for ferromagnetic instability.
However, the instability condition is not satisfied.

In multi-band models, on the other hand, it follows that
$\lim_{l \rightarrow +\infty}\lim_{n \rightarrow +0}
\bigl(\tilde{\phi}_s / \tilde{\phi}_\gamma \bigr) =1 $ for
$J=0$ while
$\lim_{l \rightarrow +\infty} \lim_{n \rightarrow +0}
\bigl(\tilde{\phi}_s / \tilde{\phi}_\gamma\bigr) > 1$ for
$J>0$.
The superexchange interaction is ferromagnetic for non-zero
$J>0$ and large enough $l$. As long as the Hund coupling is
nonzero $(J>0)$, therefore, the ferromagnetic instability
condition  Eq.~(\ref{EqFerroCond}) is generally satisfied in
the limit of low densities $(n\rightarrow+0)$, large band
multiplicity
$(l\rightarrow+\infty)$, and large spatial dimensions
$(d\rightarrow+\infty)$.\cite{ComLorentz}

\section{Discussion}\label{SecDis}

In the single-band model, the superexchange interaction is
antiferromagnetic and it  vanishes in the large limit of $U$.
The novel exchange interaction is antiferromagnetic for almost
half filling, for example, according to Fig.~\ref{Fig-A(x)}.
Therefore, the ground state must be antiferromagnetic or close
to a antiferromagnetic critical point in the thermodynamic limit of small
{\it hole} densities discussed in Introduction.  Ferromagnetic
instability cannot occur in this limit.  Nagaoka and Thouless's
ferromagnetism is a singular point as a function of carrier
numbers.


Kanamori's result\cite{Kanamori} can also be explained within
the framework of this paper. Both the flat-band and
band-edge conditions are satisfied in Kanamori's model. Then,
the novel exchange interaction is ferromagnetic so that the
ground state must be a paramagnetic state
close to a ferromagnetic quantum critical point. One can
argue that ferromagnetic instability can occur in Kanamori's
model if a weak ferromagnetic superexchange interaction is
phenomenologically included; the superexchange interaction
cannot be ferromagnetic in any single-band model.


In the flat-band models considered by Mielke and
Tasaki,\cite{Mielke1,Mielke2,Tasaki,Kusakabe} different orbits
are located at different positions within a unit cell. In
general, the Coulomb interaction between different positions is
much weaker than the on-site one. Their models correspond to
the case of $U>0$, $U^\prime=0$ and $J=0$ in Eq.~(\ref{EqInt})
within the framework of this paper. When $U^\prime$ and $J$ are
small or when $U^\prime=J$, the derivation of the superexchange
interaction in this paper breaks down, as is implied by
Eq.~(\ref{EqIij}), where its denominators vanish.  Once a large
enough but still small $U^\prime$  $(U^\prime >J)$ between
different orbits is introduced, the formulation of this paper
can be applied to such variant models of Mielke and Tasaki's.
A strong ferromagnetic superexchange interaction exists in such
models. This argument implies the existence of a strong
ferromagnetic exchange interaction or an effective strong Hund
coupling in the flat-band model of Mielke and Tasaki's, which
corresponds to the ferromagnetic superexchange interaction
considered in this paper. This is another reason why
ferromagnetism is favored in  the flat-band models.

Arguments on the flat-band condition in Sec.~\ref{SecModel} are
restricted to a rather realistic case, where a flat band lies
within broad bands. If the hybridization or mixing between flat
and broad bands is strong as it is in our model, the treatment
of this paper can be extended to the case where a flat
band lies outside broad bands. 
%
%
Because of crystalline fields in cubic lattices, for example,
$d$ orbitals split into $d\epsilon$ and $d\gamma$ multiplets. 
A model where only $d\epsilon$ or $d\gamma$ multiplet is 
taken into account corresponds to $l=1$ or $l=1/2$, while a
model where all $d$ orbitals are taken into account corresponds
to $l=2$. Although the superexchange interaction is definitely
ferromagnetic for large $l$'s, whether it is ferromagnetic or
antiferromagnetic depends rather sensitively on $l$ in the
region of small $l$'s such as $l=1$ and $l=2$, as are shown in
Figs.~\ref{FigJ(L=1*2)} and \ref{FigD(L=1*2)}. Not only orbitals
around the chemical potential but also all other orbitals far
from the chemical potential should be taken into account for a
quantitative discussion on the  magnitude of the superexchange
interaction as well as its signs. It is interesting to extend
the theory of this paper to treat such models with
non-degenerate $d$ or $f$ orbitals and models without the
symmetry of Eqs.~(\ref{EqSym}) and (\ref{EqSymU}).  Such
extensions seem to be indispensable to explain quantitatively
itinerant-electron ferromagnetism in actual metals or alloys.

It is assumed in this paper that only electrons or holes exist.
It is also interesting to extend the theory to treat
ferromagnetism in semimetals, where electrons and holes
coexist.


Because the superexchange was originally derived in insulating
phases, it is controversial whether or not it  exists in
metallic phases.  The derivation of this paper is based on
Eq.~(\ref{EqGst}) together with the Ward identity.\cite{Ward} It
is obvious that the superexchange interaction exists even in
metallic phases as long as the Mott-Hubbard splitting exists.  
When localized levels are so deep that 
$\epsilon_{a}-\mu \ll-k_{B}T_{K}$, Eq.~(\ref{EqGst}) is a good
approximation only  for
$|\varepsilon_{n}| \gg k_{B}T_{K}$. In this case, Gutzwiller's
narrow quasi-particle band appear at the chemical potential
between the lower and upper Hubbard bands.  The system is a
metal because of Gutzwiller's band. However, Gutzwiller's band
plays no role in the process of the superexchange interaction. 
When  localized levels are so shallow that
$\epsilon_{a}-\mu \agt - k_{B}T_{K}$,  Eq.~(\ref{EqGst}) is a
rather good approximation even for the vicinity of the chemical
potential; according to Eq.~(\ref{EqGst}), the mass enhancement
factor is  
$\tilde{\phi}_\gamma = 1 - n [2l/(2l+1)]$, which is 
 consistent with Gutzwiller's result given by
Eq.~(\ref{EqGutPhi}) in the large limit of
$l\rightarrow+\infty$. Gutzwiller's band and the lower Hubbard
band merge into a band and the upper Hubbard band lies above the
merged band. In both the cases, the Mott-Hubbard splitting
exists so that the superexchange interaction exists.


The superexchange interaction in metallic phases can also be
understood by the following argument. Because the typical time
scale of quasi-particles, which is $O(\hbar/k_{B}T_{B})$, is
much longer than that of pair excitations of electrons or
holes across the Mott-Hubbard gap, which is $O(\hbar/U)$ or
$O\bigl(\hbar/(U^\prime\pm J)\bigr)$, all the exchange
processes can be finished while a pair of quasi-particles meet
occasionally and part again. As long as the Mott-Hubbard
splitting exists and Eq.~(\ref{EqStrongC}) is satisfied,
therefore, the superexchange interaction works between
quasi-particle pairs that occasionally come close to each
other.


The novel exchange interaction considered in
Sec.~\ref{SecExchNov} is similar to the double and the RKKY
exchange interactions, each of which arises from that of
pair excitations of conduction electrons.  The double and the
RKKY exchange interactions are essentially the same; in many
cases the double exchange interaction seems to stand for the
${\bf q}=0$ component of the RKKY exchange interaction.
However, the novel exchange interaction is totally different
from the double or RKKY exchange interaction. When the RKKY or
double exchange interaction is derived, the existence of
localized spins are assumed. The double or RKKY exchange
interaction is an exchange interaction working between
localized spins in metals. On the other hand, the novel
exchange interaction works between itinerant
quasiparticles.  The strength of the RKKY or double
interactions is inversely proportional to the width of
conduction bands, while the strength of the novel exchange
interaction is  proportional to the bandwidth of
quasiparticles.

Because the local or intrasite part of the RKKY or double
exchange interaction can play no role, its intrasite part
should be subtracted. Then, whether the RKKY or double exchange
interaction is ferromagnetic or antiferromagnetic depends on
the position of the chemical potential. The band-edge
condition also applies to this case. When the chemical potential
is at the top or bottom region of conduction bands, it is
ferromagnetic. When the chemical potential lies in the center
region of conduction bands or the nesting of the Fermi surface
is significant, it is antiferromagnetic. The situation is
similar to that for the novel exchange interaction.

The double or RKKY exchange interaction arises from the
second-order perturbation in the $s$-$d$ exchange. When the
$s$-$d$ exchange interaction is ferromagnetic, the volume of
the Fermi surface is simply given by the number of conduction
electrons. In the band-edge model, the RKKY exchange
interaction is ferromagnetic and metallic ferromagnetism is
possible. Ferromagnetism caused by this type of the double or
RKKY exchange interaction is nothing but ferromagnetism
examined by Zener.\cite{Zener} It should never be called
itinerant-electron ferromagnetism, but it is 
local-moment ferromagnetism in metals.

When the $s$-$d$ exchange interaction is antiferromagnetic, 
localized spins must be screened by the Kondo effect.  The
analogy between the periodic antiferromagnetic $s$-$d$ model
and the periodic Anderson model implies that when higher-order
processes in the $s$-$d$ exchange interaction are considered
the double or RKKY exchange interaction should be replaced by
a similar one to the novel exchange interaction studied in
this paper.  The analogy also implies that the volume of the
Fermi surface must be given by the number of total electrons or
by the sum of those of conduction electrons and localized
spins.  When the volume of electron or hole Fermi surface is
small, this exchange interaction must be ferromagnetic. For
example, assume that there is only a single conduction band.
Denote  the average number of conduction electrons per site by
$n_{c}$. When the number of localized and screened spins is
taken into account, $n_{c}=1$ corresponds to the total
electron number 2 per site. This  exchange interaction must be
ferromagnetic for $n_{c}\simeq 1$, because the band edge
condition is satisfied. We speculate that if the critical
temperature is much lower than the Kondo temperature
ferromagnetism caused by this exchange interaction  must be
physically the same as itinerant-electron ferromagnetism
studied in this paper.\cite{ComLF}

\section{Summary}\label{SecSum}

Itinerant-electron ferromagnetism in  multi-band models is
examined within a theoretical framework of the competition
between the two leading-order effects in $1/d$: magnetic
instability caused by Weiss' mean fields or exchange
interactions and the quenching of magnetic moments by the
Kondo effect.  

Two exchange interactions are responsible for ferromagnetic
instability. One is the superexchange interaction, which arises
from the virtual exchange of pair excitations of
electrons or holes  across the Mott-Hubbard gap. When the Hund
coupling is strong enough and the band multiplicity is large
enough, the superexchange interaction becomes ferromagnetic.
The other is the novel exchange interaction arising from the
virtual exchange of pair excitations of quasi-particles, which
is also responsible for the Curie-Weiss law.  Its strength is
proportional to the bandwidth of quasi-particles, and its signs
depend on the shape of the density of quasiparticle states and
the position of the chemical potential. This exchange interaction
is ferromagnetic, when the flat-band condition is satisfied so
that the density of states has a sharp peak around the chemical
potential within or outside a broad band or broad bands or when
the band-edge condition is satisfied so that the chemical
potential is at one of the band edges.  When the electron or
hole density is close to unity per unit cell, charge
fluctuations are substantially  suppressed so that the {\it
local}\hspace{1pt} Wilson ratio is enhanced. Because spin
fluctuations are more enhanced than orbital fluctuations in the
presence of the Hund coupling, the  {\it local} Wilson ratio is
also enhanced.  The enhancement of the {\it local}\hspace{1pt}
Wilson ratio leads to the enhancement of the novel exchange
interaction. When the sum of the superexchange and the novel
exchange interactions overcomes the quenching of magnetic
moments or the suppression of magnetic instability by local and
intersite spin fluctuations, a ferromagnetic state is
stabilized at low enough temperatures.

In conclusion, the following three conditions should be
satisfied in order that a ferromagnetic state is stabilized at
low enough temperatures: (i) There are two or more than two
bands, (ii) the on-site Hubbard repulsion is much larger than
the energy scale of local quantum spin fluctuations, and (iii)
the Hund coupling is strong enough. Some or all of the
following conditions should also be satisfied: (iv) There is
an almost dispersionless band within broad ones, (v) the
chemical potential is at the top or bottom of bands,  and  (vi)
the electron or hole density is close to unity per unit cell.

\begin{acknowledgments}
This work was supported by a Grant-in-Aid for Scientific
Research (C) No.~13640342 from the Ministry of Education,
Cultures, Sports, Science and Technology of Japan.
\end{acknowledgments}

\begin{appendix}
\section{Non-mapped single-impurity  Anderson model}
\label{SecApp}

Consider ${\cal H}_{A}$ given by Eq.~(\ref{EqHamA}). In this
Appendix, it is assumed that the hybridization energy defined
by
\begin{equation}
\Delta = \frac{1}{N} \sum_{\bf k}
|v_{m{\bf k}}|^{2} \delta \bigl(\epsilon +\mu - e_{c}
({\bf k})\bigr)
\end{equation}
does not depend on  $\epsilon$ as well as $m$ and the
conduction bandwidth $2D$ is much larger than $U$, $U^\prime$
and $J$.  When $U=U^\prime=J=0$, the density of states of
localized  electrons is given by
\begin{equation}
\rho_0(\epsilon) = \frac{1}{\pi}
\frac{\Delta}{(\epsilon-\epsilon_a+\mu)^2 + \Delta^{2}}.
\end{equation}
When the Friedel sum rule or the Fermi-liquid
relation\cite{Shiba} is made use of, the density of states at
the chemical potential  for non-zero $U$, $U^\prime$ and $J$
is simply given by
\begin{equation}
\rho(0)  = \frac{1}{\pi \Delta} \sin^2
\!\left(\frac{\pi n}{2(2l+1)}\right) 
\end{equation}
as a function of  
$n = \sum_{m\sigma} \langle n_{m\sigma}\rangle$.
Our study in this Appendix is also restricted to $n <1$.

Introduce infinitesimally small external fields
given by Eq.~(\ref{EqExt}) with $H_{so}=0$, and define 
\begin{equation}
\epsilon_{m\sigma} = \epsilon_{a} 
- \Delta\mu - \frac{1}{2} \sigma g_s \mu_B   H_{s} 
- m  g_o \mu_B H_o.
\end{equation}
The ground state wave function, $\Phi$, 
is expanded in such a way that\cite{ComLargeL} 
\begin{eqnarray}\label{EqGndFunc}
\Phi &=& \Bigl[ \Gamma_0 
+ \!\sum_{m{\bf k}\sigma} 
\Gamma_{(m\sigma);({\bf k}\sigma)}
a^{\dagger}_{\sigma\tau}c_{{\bf k}\sigma} 
+ \!\sum_{{\bf k}{\bf p}\sigma}
\Gamma_{({\bf k}\sigma);({\bf p}\sigma)}
c_{{\bf k}\sigma}^{\dagger} c_{{\bf p}\sigma}
\nonumber \\  
&& \quad\quad
+  \sum_{m{\bf k}\sigma} \sum_{m^\prime{\bf p}\sigma^\prime} 
\Gamma_{[(m\sigma)(m^\prime\sigma^\prime);
({\bf p}\sigma^\prime)({\bf k}\sigma)]}\times
\nonumber \\ &&  \hspace*{2.5cm} 
\times 
a^{\dagger}_{m\sigma}a^{\dagger}_{m^\prime\sigma^\prime}
c_{{\bf p}\sigma^\prime}c_{{\bf k}\sigma}
%
+ \cdots \Bigr] |0\rangle ,
\end{eqnarray}
with $|0\rangle$ the Fermi vacuum, and it satisfies 
\begin{equation}\label{EqSchA}
\left[E - {\cal H}_{A}-{\cal H}_{ext} \right] \Phi = 0 .
\end{equation}
In Eq.~(\ref{EqGndFunc}), 
$\Gamma_{[(m\sigma)(m^\prime\sigma^\prime);
({\bf p}\sigma^\prime)({\bf k}\sigma)]}$ \
is an antisymmetrized coefficient defined in such a way that
\begin{eqnarray}
\Gamma_{[(m\sigma)(m^\prime\sigma^\prime);
({\bf p}\sigma^\prime)({\bf k}\sigma)]} &=& \nonumber \\ 
&& \hspace*{-3.1cm} = 
\Gamma_{(m\sigma)(m^\prime\sigma^\prime);
({\bf p}\sigma^\prime)({\bf k}\sigma)} \!-\!
\Gamma_{(m^\prime\sigma^\prime)(m\sigma);
({\bf p}\sigma^\prime)({\bf k}\sigma)} \nonumber \\  
&& \hspace*{-3.0cm} -
\Gamma_{(m\sigma)(m^\prime\sigma^\prime);
({\bf k}\sigma)({\bf p}\sigma^\prime)} \!+\!
\Gamma_{(m^\prime\sigma^\prime)(m\sigma);
({\bf k}\sigma)({\bf p}\sigma^\prime)} .
\end{eqnarray}
Eqs.~(\ref{EqGndFunc}) and (\ref{EqSchA}) give a set
of equations for  $\Gamma$'s:
\begin{mathletters}\label{EqAs}
\begin{equation}\label{EqA0}
E \Gamma_0 - \frac{1}{\sqrt{N}}
\sum_{m{\bf k}\sigma} v_{{\bf k}\tau}^{*}
\Gamma_{(m\sigma);({\bf k}\sigma)} =0 ,
\end{equation}
\begin{eqnarray}\label{EqA1}
&&
\Bigl[ E -\epsilon_{m\sigma}+ e_c ({\bf k}) \Bigr]
\Gamma_{(m\sigma);({\bf k}\sigma)}
- \frac{1}{\sqrt{N}} v_{m{\bf k}} \Gamma_0 -
\nonumber \\ && \quad 
- \frac{1}{\sqrt{N}} \sum_{{\bf p}} v_{m{\bf p}}
\Gamma_{({\bf p}\sigma);({\bf k}\sigma)}
\nonumber \\ && \quad 
- \frac{2}{\sqrt{N}} 
\sum_{m^\prime{\bf p}\sigma^\prime} v_{m^\prime{\bf p}}^{*}
\Gamma_{[(m\sigma)(m^\prime\sigma^\prime);
({\bf p}\sigma^\prime)({\bf k}\sigma)]} =0 ,
\end{eqnarray}
\begin{eqnarray}\label{EqA2}
&&\Bigl[E - e_{c}({\bf p}) + e_{c}({\bf k}) \Bigr]
\Gamma_{({\bf p}\sigma);({\bf k}\sigma)} +
\nonumber \\ && \hspace*{2.0cm}
- \frac{v_{m{\bf p}}^{*}}{\sqrt{N}} 
\Gamma_{(m\sigma);({\bf k}\sigma)} + \cdots =0 ,
\end{eqnarray}
\begin{equation}\label{EqA30} 
\Gamma_{[(m\sigma)(m\sigma);
({\bf p}\sigma)({\bf k}\sigma)]} =0 , 
\end{equation}
\begin{eqnarray}\label{EqA31} 
&& \Bigl[ 
 E -\epsilon_{m\sigma} -\epsilon_{m-\sigma} - U +
\nonumber \\ && \hspace*{1.0cm}
+ e_c ({\bf k})  + e_{c}({\bf p}) \Bigr] 
\Gamma_{[(m\sigma)(m-\sigma);
({\bf p}-\sigma)({\bf k}\sigma)]} \ -
\nonumber \\ && \hspace*{0.5cm} 
- \frac{v_{m{\bf p}}}{\sqrt{N}}
 \Gamma_{(m\sigma);({\bf k}\sigma)}
- \frac{v_{m{\bf k}}}{\sqrt{N}}
 \Gamma_{(m-\sigma);({\bf p}-\sigma)} + \cdots =0 , 
\nonumber \\
\end{eqnarray}
\begin{eqnarray}\label{EqA32} 
&&\Bigl[  E -\epsilon_{m\sigma} -\epsilon_{m^\prime\sigma} -
U^\prime + J +
\nonumber \\ && \hspace*{2.0cm}
 + e_c ({\bf k}) + e_{c}({\bf p})
\Bigr] 
\Gamma_{[(m\sigma)(m^\prime\sigma);
({\bf p}\sigma)({\bf k}\sigma)]} -
\nonumber \\ && \hspace*{0.5cm} 
- \frac{v_{m^\prime{\bf p}}}{\sqrt{N}} 
\Gamma_{(m\sigma);({\bf k}\sigma)}
- \frac{v_{m{\bf k}}}{\sqrt{N}}
 \Gamma_{(m^\prime\sigma);({\bf p}\sigma)}  + \cdots =0 
\nonumber \\
\end{eqnarray}
\begin{eqnarray}\label{EqA33} 
&& \Bigl[E- \epsilon_{m\sigma} - \epsilon_{m^\prime-\sigma} -
U^\prime +
\nonumber \\ && \hspace*{1.5cm} + 
e_{c}({\bf k}) +e_{c}({\bf p})
\Bigr] 
\Gamma_{[(m\sigma)(m^\prime-\sigma);
({\bf p}-\sigma)({\bf k}\sigma)]} +
\nonumber \\ && \hspace*{0.5cm}  
+ J \Gamma_{[(m-\sigma)(m^\prime\sigma);
({\bf p}-\sigma)({\bf k}\sigma)]}  \nonumber \\
&& \hspace*{0.5cm} 
- \frac{v_{m^\prime{\bf p}}}{\sqrt{N}} 
\Gamma_{(m\sigma);({\bf k}\sigma)}
\!-\! \frac{v_{m{\bf k}}}{\sqrt{N}}
 \Gamma_{(m^\prime-\sigma);({\bf p}-\sigma)}  + \cdots =0 ,
\!\! \phantom{\Bigg]} \nonumber \\
\end{eqnarray}
\begin{eqnarray}\label{EqA34} 
&&
\Bigl[ E- \epsilon_{m-\sigma} - \epsilon_{m^\prime\sigma}  -
U^\prime +
\nonumber \\ && \hspace*{1.5cm}
  +e_{c}({\bf k})+ e_{c}({\bf p})
\Bigr] 
\Gamma_{[(m-\sigma)(m^\prime\sigma);
({\bf p}-\sigma)({\bf k}\sigma)]} +
\nonumber \\ && \hspace*{1cm} 
+ J \Gamma_{[(m\sigma)(m^\prime-\sigma);
({\bf p}-\sigma)({\bf k}\sigma)]}
 + \cdots =0 ,\phantom{\Bigr]}
\end{eqnarray}
\end{mathletters}
and so on. In Eqs.~(\ref{EqA32}), (\ref{EqA33}) and
(\ref{EqA34}), $m \ne m^\prime$.

When  small irrelevant terms are ignored, 
it follows from Eqs.~(\ref{EqA0})--(\ref{EqA34}) that
\begin{equation}
\frac{1}{F(E)} =
\frac{\Delta}{\pi} \sum_{m\sigma} \ln \left|
\frac{D}{E- \Delta E_{0} -\epsilon_{m\sigma}^{*} 
+\mu} \right| ,
\end{equation}
with
\begin{eqnarray}
F(E) &=& 
\frac{1}{-E}
+ \frac{1}{2l+1} \Biggl[
\frac{1}{U + \epsilon_{a} - \mu} +
\nonumber \\
&& 
+ \frac{3l}{ U^\prime - J + \epsilon_{a} - \mu}
+ \frac{l}{U^\prime + J + \epsilon_{a} - \mu}
\Biggr] , 
\end{eqnarray}
\begin{eqnarray}
\Delta E_{0} &=& - \frac{\Delta}{\pi} \Biggl[ 
\ln \frac{D}{\Delta} + 2\ln \frac{D+U}{U}
\nonumber \\ && 
+ 3l  \ln \frac{D+U^\prime-J}{U^\prime - J}
 + l \ln \frac{D+U^\prime+J}{U^\prime + J}
\Biggr] ,
\end{eqnarray}
and
\begin{equation}\label{EqEST}
\epsilon_{m\sigma}^{*} = \epsilon_{a} - g_c^* \Delta\mu
- \frac{1}{2}\sigma g_{s}^* \mu_B H_{s} - m g_o^* \mu_B H_o .
\end{equation}
Here, $g_c^*$, $g_s^*$ and $g_o^*$ are 
renormalized $g$ factors: 
\begin{equation}
g_c^* = 1 \!+\! \frac{1}{2l\!+\!1} \! \bigl[ R(U) 
\!+\! 3l R(U^\prime\!-\!J)+  l R(U^\prime\!+\!J) \bigr],
\end{equation}
\begin{equation}
\frac{g_s^*}{g_s} = 1 + \frac{2l}{2l+1} R(U^\prime-J),
\end{equation}
and
\begin{eqnarray}\label{EqGo}
\frac{g_o^*}{g_o} &=& 1+\frac{1}{2l+1}R(U) 
\nonumber \\ && 
 + \frac{2l-1}{2l+1}\left[ \frac{3}{4} R(U^\prime-J)  
+ \frac{1}{4} R(U^\prime+J) \right] ,
\end{eqnarray}
with 
\begin{equation}
R(x) = \frac{4(2l\!+\!1)\Delta}{\pi} \!\left[
\frac{1}{x\!+\!\epsilon_{a}\!-\!\mu}
-\frac{1}{D\!+\!x\!+\!\epsilon_{a}\!-\!\mu}
\right].
\end{equation}
Eq.~(\ref{EqGo}) should be discarded for $l=0$. Note that 
$g_c^* > g_s^*/g_s \ge g_o^*/g_o > 1$;
$g_s^*/g_s=g_o^*/g_o$ when
$U=U^\prime$ and $J=0$, while $g_s^*/g_s> g_o^*/g_o$ when
$U>U^\prime$ or $J>0$. Even in the absence of the spin-orbit
interaction, the $g$ factor in the spin channel is enhanced in
multi-band models.

When $\langle n_{m\sigma} \rangle \ll 1$,
$F(E)$ is approximated by $F(E) \simeq - 1/E$.
%
%
%
The electron number is given by
$n = - \partial E/\partial \Delta\mu$. 
The charge susceptibility is calculated so that 
\begin{eqnarray}
\tilde{\chi}_{c}^* (0) &=&
\frac{\partial n\phantom{\Delta}}{\partial \Delta\mu}
= \left(g_c^*\right)^2 
\!\!\frac{\pi}{2(2l\!+\!1)\Delta}
\hspace{1pt}n^{2} \left(g_c^* -n\right)
\end{eqnarray} 
as a function of $n$. In similar manners, spin and orbital
susceptibilities are calculated so that
\begin{eqnarray}
\tilde{\chi}_{\alpha}^*(0) &=& 
\left(\frac{g_\alpha^*}{g_\alpha}\right)^{2}
\hspace{-5pt} \frac{\pi}{2(2l\!+\!1)\Delta}
\hspace{1pt}\frac{n^{2}}{g_c^*-n} .
\end{eqnarray} 
The expansion coefficients appearing in the single-site 
selfenergy are given by
\begin{equation}
\tilde{\phi}_{c} = (g_c^*)^2 (g_c^*-n) \hspace{1pt}
\frac{\langle \pi n_{m\sigma}
\rangle^2}  {\sin^2\langle \pi n_{m\sigma} \rangle}
\end{equation}
for the charge channel, and
\begin{equation}
\tilde{\phi}_{\alpha} =
\left(\frac{g_\alpha^*}{g_\alpha}\right)^{2}
\frac{1}{g_c^*-n} \hspace{1pt}
\frac{\langle \pi n_{m\sigma} \rangle^2} 
{\sin^2\langle \pi n_{m\sigma} \rangle}
\end{equation}
for spin $(\alpha=s)$ and orbital $(\alpha=o)$ channels. 
Note that
%
$\tilde{\phi}_c > \tilde{\phi}_s \ge \tilde{\phi}_o > 1$
%
in the low density limit of $n \rightarrow +0$  although
%
$\tilde{\phi}_s \ge \tilde{\phi}_o \gg 1 \gg \tilde{\phi}_c$ 
%
for $n \simeq 1$.
\end{appendix}


\begin{references}
\bibitem{ComCW}
The Curie-Weiss law of local-moment magnets arises from the
temperature dependence of local thermal spin fluctuations.  In
magnets, there are only two relevant leading order effects in
$1/d$: local spin fluctuations and magnetic Weiss' mean fields. 
Local charge fluctuations and isotropic BCS superconducting
fluctuations and their corresponding Weiss' mean fields are also
of leading order in $1/d$, but they are irrelevant in magnets.
\bibitem{Com1}
When no interband terms are included, the formulation of this
paper can be applied to single-band models.
\bibitem{Com2} 
This relation can be only approximately satisfied in actual
systems.
\bibitem{ComValence}
In this case, the mapped Anderson model lies in the so called
valence fluctuation regime. The Kondo temperature or energy
can be still defined as the energy scale of local quantum spin
fluctuations by Eq.~(\ref{EqDefTK}).
\bibitem{ComSubtract}
Intraband contributions $(r=r^\prime)$ are never significant,
unless they are in the vicinity of the chemical potential.
\bibitem{ComLorentz}
When $\rho^*(x)$ is of a Lorentzian shape in such a way that
$\rho^*(x) = \left(\Delta^*/\pi\right)/
\left[\left(x-\epsilon_a^*\right)^2 +
\left(\Delta^*\right)^2\right]$, 
the novel exchange interaction vanishes for any
$\epsilon_a^*$ at $T=0$~K because  
%
\[
\int_{-\infty}^{+\infty} \hspace{-10pt} dx 
\int_{-\infty}^{+\infty} \hspace{-10pt} dy
\frac{f(x)-f(y)}{y-x} \rho^*(x) \rho^*(y) 
= \rho^* (0) .
\]
\bibitem{ComLF}
If $T\gg T_K$, localized spins are not screened. Then, the
double or RKKY exchange interaction must be relevant for
$T \gg T_K$, and it is presumably antiferromagnetic because the
volume of the Fermi surface is large for $n_c \simeq 1$.  A
ferromagnetic state can never be stabilized above $T_K$ for
$n_c\simeq 1$.
\end{references}
\end{document}